\documentclass[%
nofootinbib,
nobibnotes,
amsmath,amssymb,
aps,
twocolumn,
prb]{revtex4-2}

\usepackage{lineno}
\usepackage{amsmath}
\usepackage{amssymb}
\usepackage{overpic}
\usepackage{amsfonts}
\usepackage{graphicx}
\usepackage{dsfont}
\usepackage{bbold}
\usepackage{bm}
\graphicspath{{./imgs/}{./imgs_1/}}
\usepackage{xcolor}
\usepackage{ifthen}
\usepackage{braket}
\usepackage{hyperref}
\usepackage[caption=false, subrefformat=parens]{subfig}

\usepackage[capitalize,compress]{cleveref}
\crefname{section}{Section}{Section} %
\crefname{subsection}{Section}{Section}
\crefname{equation}{Eq.}{Eqs.}
\creflabelformat{equation}{#2#1#3} %
\Crefname{equation}{Equation}{Equations}

\newcommand{\tr}[1][]{\ifthenelse{\equal{#1}{}}{\mathrm{Tr}\,}{\mathrm{Tr}\left[#1\right]}}

\newcommand{\vb}[1]{#1} %
\newcommand{\wc}{w_\mathrm{c}}
\newcommand{\omegac}{\omega_\mathrm{c}}

\usepackage{physics}

\begin{document}

\title{%
Integer Quantum Hall Effect: Disorder, temperature, localization, floating, and plateau width%
}

\author{Stuart Yi-Thomas}
\email[]{snthomas@umd.edu}
\affiliation{Condensed Matter Theory Center and Joint Quantum Institute, University of Maryland, College Park, Maryland 20742-4111, USA}
\author{Yi Huang}
\affiliation{Condensed Matter Theory Center and Joint Quantum Institute, University of Maryland, College Park, Maryland 20742-4111, USA}
\author{Jay D. Sau}
\affiliation{Condensed Matter Theory Center and Joint Quantum Institute, University of Maryland, College Park, Maryland 20742-4111, USA}
\author{Sankar Das Sarma}
\affiliation{Condensed Matter Theory Center and Joint Quantum Institute, University of Maryland, College Park, Maryland 20742-4111, USA}

\date{January 14, 2025}

\begin{abstract}

We theoretically consider disorder and temperature effects on the integer quantum Hall effect (IQHE) using a variety of distinct and complementary analytical and numerical techniques.  In particular, we address simple, physical, and experimentally relevant questions: How does disorder and/or temperature affect the IQHE plateau width? Does the plateau width increase or decrease with disorder and/or temperature? What happens to the peak in the longitudinal conductance with increasing disorder/temperature?  Does the longitudinal conductance obey any universal scaling property? Is there ``floating'' with increasing disorder and/or decreasing magnetic field? Can disorder destroy the IQHE? Is there an IQHE to localization transition?  What is the Landau level dependence of the plateau width?  Our detailed theory provides answers to these and other related experimentally relevant questions.  We discuss our results in the context of existing experimental results and suggest future experiments arising from our work.  A key finding is that disorder and temperature are intrinsically connected in affecting IQHE, and there is an intricate interplay between them leading to nonmonotonicity in how the IQHE plateau width behaves as a function of increasing disorder. Both must be considered on an equal footing in understanding IQHE experiments.
\end{abstract}

\maketitle
\section{Introduction}
The integer quantum Hall effect (IQHE), serendipitously discovered in 1980 \cite{vonklitzing1980}, is among the most profound and important phenomena in all of physics.  On a practical level, IQHE provides the resistance standard because of the precise quantization of the Hall resistance in the quantum Hall plateaus~\cite{codata}, and thus helps define the fundamental constants $e$ and $h$~\cite{cgpm}. On a fundamental level,  IQHE is also the first experimental discovery of a topological phenomenon in physics as the quantization is thought to arise from the intrinsic existence of a Chern number underlying IQHE.  As such, IQHE is the first reported ``topological phase'' where the quantum phase is defined through a topological index (``the Chern number'') and not by an order parameter.  IQHE is a well-established phenomenon and has been extensively reviewed in the literature~\cite{prange1989,dassarma1997,yoshioka2013,Janssen:1994,tong2016,vonklitzing2017,witten2016}.

In spite of an enormous amount of work and an apparent consensus that IQHE is well-understood (because it is basically a manifestation of topological invariance when the Fermi level is in a bulk gap between Landau levels \cite{thouless1982,avron1983,niu1985,Thouless1981,streda1982}), many specific questions remain open as they are outside the scope of the topological invariance paradigm which only applies for infinite systems at $T=0$.  
For example, while the quantization at $T=0$ is exact in the thermodynamic limit as long as the longitudinal resistance ($\rho_{xx}$) vanishes---because it is a Chern number (in units of $h/e^2$ for the Hall resistance, $\rho_{xy}$)---the correction to the quantization at finite temperatures is unknown since there is no microscopic transport theory for IQHE and we cannot predict the value of $\rho_{xy}$ when $\rho_{xx}$ is nonzero (as it always is at finite temperatures even if it may be exponentially suppressed for temperatures much lower than the inter-Landau level energy gap).  
The Chern number invariance and the closely related Chern-Simons field theory for IQHE \cite{witten2016} provide no hint on how to calculate these finite temperature corrections since they are necessarily zero temperature theories.
The same theoretical ignorance also applies to the very physical question of what happens to a particular IQHE plateau when disorder increases.
(Again, the Chern-Simons field theory for IQHE, while establishing the exactness of the quantization, is a zero parameter theory which says nothing whatsoever about the plateau.)
We do not know theoretically whether the plateau expands or shrinks with increasing disorder!  
Experiments show that plateau may either expand~\cite{Klitzing:1985,Furneaux:1984,Furneaux:1986,Gottwaldt:2003,mohle:1989465,Adrian:1989,Sigg:1988293,Stormer:198232,Harmand:2009} or shrink~\cite{Furneaux:1986} with increasing disorder.
A fundamental question receiving early attention is the so-called ``floating''~\cite{khmelnitskii1984,laughlin1984,fogler1998quasiclassical}, where the extended states at each Landau level center float up in energy as the magnetic field decreases or disorder increases.
(This implies that as the dimensionless quantity $\omegac \tau$ decreases, the quantized Landau level structure becomes increasingly weaker, where $\omegac$ is the cyclotron frequency and $\tau$ is the transport relaxation time.)  
The floating hypothesis enables a qualitative reconciliation between the essential existence of extended states (at the center of each Landau level where the longitudinal conductivity is nonzero) in the presence of a magnetic field with the known fact that the zero field 2D system must be an Anderson localized insulator in the presence of any disorder because of the complete destructive interference caused by 2D backscattering. 
See, e.g., the article by Das Sarma~\cite{dassarma1997}.
This hypothesis conjectures that the extended states at Landau level centers must levitate to infinite energy as the magnetic field vanishes (or equivalently as the $\omega_c \tau$ parameter vanishes)~\cite{glozman1995}. Floating, while seemingly reasonable in reconciling IQHE with the known weak localization physics in 2D systems, is rarely demonstrated explicitly in continuum theoretical calculations~\cite{sheng2000,sheng2001}.

A closely related question is what happens when the disorder is very strong, in particular, whether a very strong disorder by itself can destroy IQHE in the strong-field Landau level situation for a critical disorder strength, leading to localization directly without going through any floating. We note that these questions remain open with no clear answers, particularly for the experimentally relevant systems, in spite of occasional efforts to address them~\cite{levine1983,pruisken1984,kivelson1992,glozman1995,sheng2000,sheng2001,Hannahs:1993,Kirk:1994,wong1995,Monroe:1997,Pudalov:1993,Pudalov:1994,Pudalov:1990,XC_Xie:1996,Pudalov:1995,huo1993,shahar1995}.  We mention that the alternative (but equivalent to the Chern number paradigm of IQHE) edge state picture of IQHE, where the quantization arises essentially from the discreteness of electrons (i.e.\ the conservation of electron numbers as natural integers), also allows no methodologies to address these questions involving temperature and disorder effects on IQHE~\cite{laughlin1984,halperin1982}.
The abstract theories involving Chern numbers, Chern-Simons theory or edge considerations provide elegant general arguments for the $T=0$ quantization (assuming implicitly that the chemical potential is pinned deep inside a mobility gap between consecutive Landau levels), but tell us nothing about the deviations from the exact quantization arising necessarily in real samples. See, for example, Ref.~\onlinecite{prange1989}.

In our theory, electron localization arises either from destructive quantum interference between multiple-scattering paths induced by disorder (Anderson localization~\cite{anderson1958,ioffe1960}) or from classical localization resulting from the interplay of magnetic fields and disorder. The latter includes cyclotron orbits drifting along closed equipotential lines in a long-range disordered potential landscape~\cite{Laikhtman:1994,Fogler:1997} and electrons following rosette-like trajectories around short-range impurities~\cite{Baskin:1978}. Collectively, we refer to these phenomena as disorder-induced localization. In the thermodynamic limit, each Landau level hosts a single delocalized state responsible for a longitudinal conductance peak as a function of the filling factor; the plateau transition occurs precisely at this point. Experimental observations of this localization transition within a Landau level are established through measuring the width of the conductance peak vs.\ sample size~\cite{Klitzing:1992}. Accordingly, we use the longitudinal conductance peaks calculated in our model to determine the positions of the delocalized states and the corresponding plateau widths. This correspondance was also found in Ref.~\onlinecite{dresselhaus2022}.

One crucial experimental parameter is the Fermi energy or the chemical potential, which does not appear directly in the Hamiltonian, but controls the number of occupied Landau levels.  This is an additional energy scale which plays a key role as it makes a huge difference whether the chemical potential is in the lowest Landau level or in some high Landau levels as the lowest Landau level has a special role because it does not have any occupied level below it, and eventually IQHE must disappear at some (perhaps low) filling of the lowest Landau level.  
Experimentally, the magnetic field is often (but not always) varied to study IQHE, keeping the chemical potential (controlled by the 2D electron density) constant.  Increasing/decreasing the applied magnetic field makes consecutive Landau levels go through the Fermi level, but this is inconvenient from the theoretical perspective since changing the magnetic field changes the Hamiltonian (the effective temperature $T/\hbar \omega_c$ also changes if $T\neq0$).  We therefore always consider the constant magnetic field situation so that both the cyclotron energy and the magnetic length are constant. 
We assume that the electron density and therefore, the chemical potential, is being tuned in the system to move through the Landau levels, starting from the lowest Landau level (corresponding to zero electron density) upward with all the Landau levels having equal separation as the Fermi level sweeps through each Landau level starting from the lowest one.  This Fermi energy introduces an additional energy scale in the problem defining the number of occupied levels.
We note that of the four important experimental parameters, two (namely, cyclotron energy and disorder) enter the Hamiltonian while the other two (Fermi energy and temperature) do not.

In the current work, we address these questions using theoretical techniques, incorporating both finite disorder and finite temperature in the theory.  
We use several distinct but complementary methods to develop a comprehensive picture of IQHE going beyond the Chern number paradigm which cannot address any of the nonuniversal questions relevant to the experiments. 
Our findings, to be described in depth in the next sections of this article, are nuanced and intricate because of the complications of multiple independent length and energy scales in the problem.

The relevant energy scales are cyclotron energy (defining the Landau level gaps), the disorder strength (defining the Landau level broadening and also localization), and temperature (defining level occupancy).  
The corresponding length scales are Landau radius or magnetic length, disorder correlation length, thermal length, and the system size---in some sense the finite system size may act as an effective localization length scale or as a temperature itself through the concept of a cut off of the coherence.  
In our theory, we do not assume any of these energy scales dominate, finding that the dimensionless disorder/temperature, disorder/cyclotron energy, and cyclotron energy/temperature may all play crucial roles depending on the details.
Given the complexity of the problem, we make several simplifying approximations, all of which are routine in IQHE theories.  
We assume noninteracting electrons and parabolic/isotropic 2D systems with a magnetic field oriented normal to the 2D layer with no spin-orbit coupling.  
We neglect spin (and any valley) degeneracy.  
These are all non-essential approximations (except for the neglect of interactions), and we expect our qualitative conclusions to remain unaffected by these assumptions.  
Finally, we use several different specific (and complementary) microscopic models, which are described in detail in each section below where they are discussed and their results presented to avoid repetitions.
Below we provide a synopsis of each section.

In Sec.~\ref{sec:tight-binding}, we use a microscopic tight-binding model to study the effect of correlated disorder on the IQHE plateaus. We find using an exact conductance calculation that at zero temperature, the longitudinal conductance peaks (arising from the extended states at the center of each Landau level) float up to higher fillings and the spacing between them decreases as disorder increases, indicating that the plateaus shrink with increasing disorder at $T=0$. 

In Sec.~\ref{sec:percolation_model}, we study how temperature and disorder affect a percolation model of IQHE where the disorder spatially varies slowly over the magnetic length scale.
At a finite temperature, the plateau width initially increases with disorder for small disorder strengths but eventually narrows at higher disorder due to floating, being in qualitative agreement with the $T=0$ results of Sec.~\ref{sec:tight-binding}.
In addition, we find a classical percolation transition from IQHE to an insulating state where electrons are separated into inhomogeneous puddles when the disorder becomes larger than the Fermi energy (which can be viewed as the generalized Anderson-Ioffe-Regel criterion~\cite{anderson1958,ioffe1960} in the presence of a magnetic field). 
All floating physics occurs for a disorder smaller than the Fermi energy.

We conclude in Sec.~\ref{sec:conclusion} by summarizing our findings and their implications for experiments, and also discuss any remaining open questions. An appendix provides some additional numerical results for completeness.

\section{Tight-binding model}\label{sec:tight-binding}
We begin by describing the disordered IQHE system using a microscopic tight-binding model at zero temperature. By using the open source scattering matrix software package \textsc{kwant}~\cite{kwant} to exactly solve the scattering problem, we compute both the longitudinal conductance and the filling in the presence of arbitrary disorder. The peaks in the longitudinal conductance correspond to boundaries between quantized Hall conductance plateaus and allow inference about the width of each plateau with varying disorder. 

Note that some level of infinitesimal disorder is necessary to produce the IQHE since the chemical potential must reside in an inter-Landau-level spectral gap, corresponding to a mobility gap of localized states, to produce a Hall conductance plateau.  Without any disorder, the Fermi energy is restricted to the exact energies of the extended states which form a set with zero measure. However with an infinitesimal value of disorder, the plateaus are perfect at $T=0$, covering the entire gap regime between Landau levels.

\begin{figure}
    \centering \includegraphics[width=\linewidth]{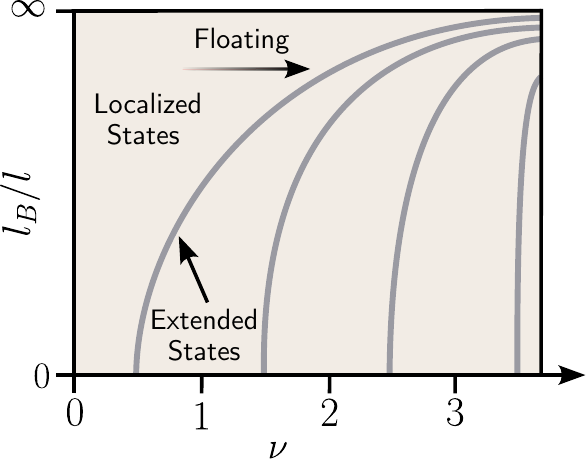}
    \caption{%
    Schematic for the zero temperature phase diagram demonstrating floating as the disorder-induced mean free path $l$ decreases relative to the magnetic length $l_B$. At low disorder, the Landau levels are located at half integers and float to higher fillings with increasing disorder. Note that we consider a fixed magnetic field keeping the magnetic length constant, and change the chemical potential by increasing density.  Note also that the floating or the levitation of the individual Landau levels slows down with increasing Landau level number, and the disorder-induced localization due to floating begins at the lowest Landau level, moving up in Landau level as disorder increases.%
    \label{fig:phase_diagram}}
\end{figure}

Using zero-temperature simulations at various values of disorder, we find in Sec.~\ref{sec:floating} that the plateau width decreases with increasing disorder as the conductance peaks float up to higher fillings (see Fig.~\ref{fig:phase_diagram}). 
The result is disorder-induced localization at fillings lower than the lowest Landau level (i.e.\ filling less than unity). 
Importantly, this conclusion is only valid for zero temperature. When we extend the results to finite temperature in Sec.~\ref{sec:finite-temperature}, we find that the plateaus can indeed expand depending on the relative magnitudes of temperature and disorder.
To justify these conclusions, we analyze the finite-size scaling of the longitudinal conductance peaks in Sec~\ref{sec:peak-scaling} and find a rough scaling relation consistent with previous works~\cite{dresselhaus2022, huckestein1995, Polyakov:1993, polyakov1993b}, however the mixing of Landau levels precludes a universal relationship.

We calculate the longitudinal conductance for a disordered 2D electron gas (2DEG) in a magnetic field within the tight-binding approximation described by the Hamiltonian
\begin{equation}
H = -\sum_{\langle i j \rangle} t \,e^{ia_{ij}}\, c_i^\dagger c_j + \mathrm{h.c.} + \sum_i V_i c_i^\dagger c_i
\end{equation}
with periodic boundary conditions in the transverse direction. (Our geometry is thus a cylinder.)
We use the Landau gauge which manifests as a Peierls phase of $a_{ij} = x_i (y_j - y_i)/l_B^2$ in the kinetic energy where $l_B=\sqrt{\hbar / e B}$ is the magnetic length, $B$ is the magnetic field, $e$ is the electron charge and $x_i$ and $y_i$ are the position coordinates of the $i$-th lattice site. 
We take the disorder $V_i$ to be spatially correlated over a length scale $d$ and to have strength $w$ such that variance over realizations is $\langle V_i V_j \rangle = w^2 \exp(-|r_i-r_j|^2/2 d^2)$ where $d=2$ in units of the lattice constant. 
We predominantly use a magnetic length of $l_B=4$ in terms of the lattice constant $a$, which we take to be unity. From the tight-binding hopping $t=\hbar^2/2m^* a^2$ where $m^*$ is the effective electron mass of the electrons in the 2DEG, we derive the cyclotron frequency $\omegac = 2t (a/l_B)^2/\hbar$ which is $0.126$ where we set $\hbar$ and $t$ to unity. Note that this expression for the cyclotron frequency assumes a parabolic dispersion and negligible lattice effects. The former is valid since we consider energies on the order of $\omegac$ which is much smaller than the hopping and the magnetic length is longer than the lattice constant.
We also mention that our choice of $d=2$ and $l_B=4$ are not restrictive, and the qualitative results do not change for other choices.  The choice of $d<l_B$ makes the disorder short-ranged which is more applicable to strongly disordered 2D systems.  (In Sec.~\ref{sec:percolation_model} we consider the long-range disorder regime of $d>l_B$, where a percolation model is more appropriate.)

We use \textsc{kwant}~\cite{kwant} to calculate both the conductance and the filling. The conductance is calculated exactly by attaching full-width leads to a square sample of size $L\times L$ and solving for the transmission problem in the scattering region. We calculate the conductance for multiple disorder realizations and approximate the average using the median since the distribution is skewed~\cite{dresselhaus2022}. (As can be seen in Appendix~\ref{sec:kwant_app} the distribution is essentially log normal.) The filling is given by the cumulative density of states which is calculated using the kernel polynomial method~\cite{kwant,kpm}. We use 100 random vectors and interpolate to a polynomial with 1000 moments. Such a large number is necessary to include quantum oscillations in the density of states. The system sizes we use (up to $L=800$) are at the limits of our computational resource, but we do not believe that going to larger systems would lead to any new qualitative insights. We provide details regarding these calculations in Appendix~\ref{sec:kwant_app}.

\subsection{Floating}
\label{sec:floating}
\begin{figure}[t]
    \centering
    \includegraphics[width=\linewidth]{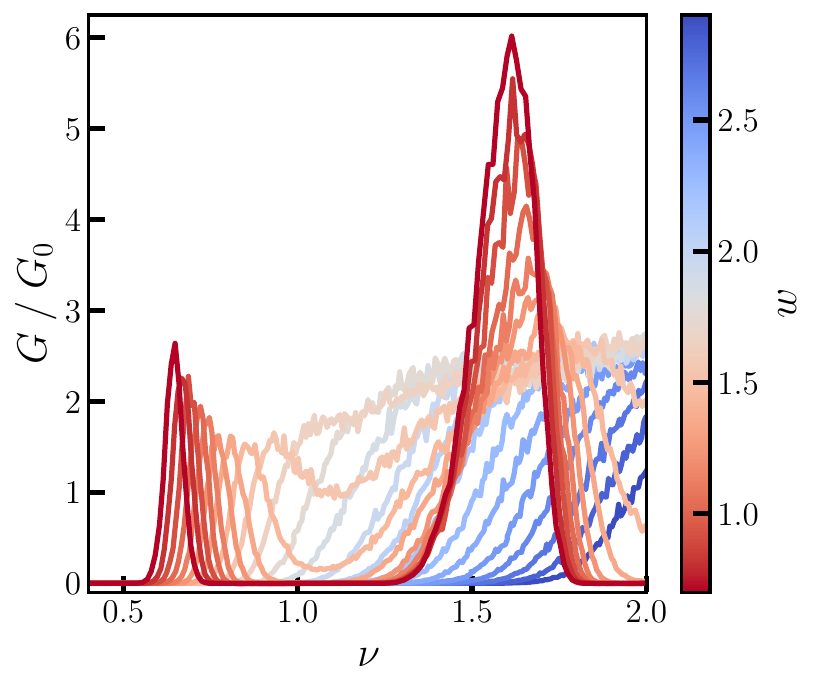}
    \caption{%
    Longitudinal conductance $G$ (scaled by $G_0=e^2/h$) of the first two Landau levels as a function of filling $\nu$ with increasing disorder strength $w$.
    $L=800$, $l_B=4$. We use approximately $1200$ disorder realizations.  %
    \label{fig:floating}}
\end{figure}

The longitudinal conductance $G$ as a function of the filling factor $\nu$, resulting from the above calculation, is shown in Fig.~\ref{fig:floating}.
We can identify the Landau levels in this plot from the peaks in the conductance which indicate the extended states at the Landau level centers.
At zero temperature, increasing disorder causes Landau levels to float up to higher filling factors, as is obvious from the rightward shift of the peaks to higher $\nu$ values with increasing disorder. 
From these conductance peaks, we infer that the conductance plateaus in the Hall conductance diminish as the longitudinal conductance peaks rise to higher and higher fillings since the spacing between peaks decreases with increasing disorder in Fig.~\ref{fig:floating}, thus providing direct evidence for floating. The result is an expansion of the localized state beneath the lowest Landau level arising from the tail in the density of states extending into negative energies (i.e.\ localized) below the lowest Landau band.  

To measure the shrinkage of the plateau between conductance peaks, we introduce two parameters: $\Delta \nu_1$, the distance between the peaks and $\Delta \nu_2$, the size of the mobility gap. 
Assuming that the conductance peaks narrow to a set of measure zero (see Sec.~\ref{sec:peak-scaling}), as expected in the thermodynamic limit, these two metrics should quantitatively converge at infinite system size.
They behave qualitatively similarly in Fig.~\ref{fig:floating} although the separation between the peaks is somewhat larger than the mobility gap as the calculated conductance peaks are not infinitely sharp because of finite size effects.

\begin{figure}[t]
\centering
    \begin{overpic}[width=1\linewidth]{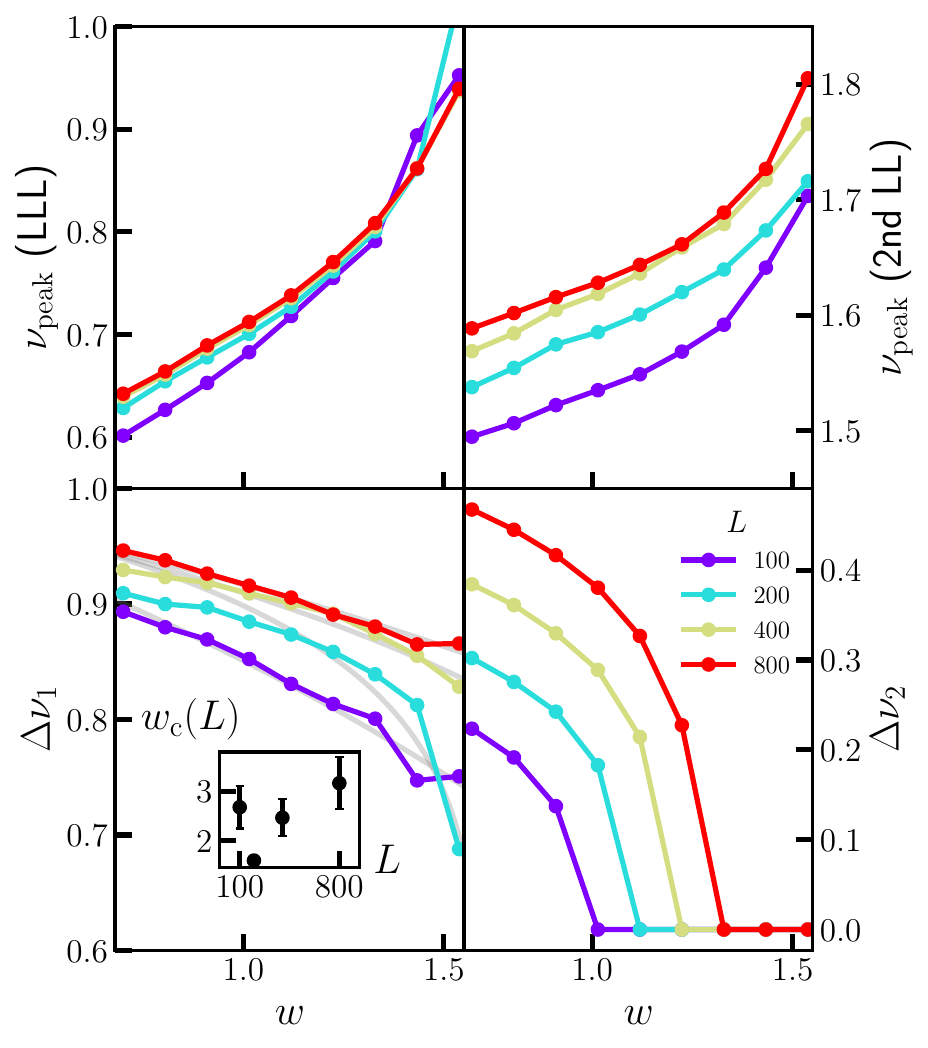}
    \put(14,56){(a)}
    \put(14,12){(b)}
    \put(47,56){(c)}
    \put(47,12){(d)}
    \end{overpic}
    \caption{\label{fig:peak-positions} Finite-size scaling of conductance peaks corresponding to the lowest two Landau levels. (a,c) Position of Landau levels. (b) Distance between peaks $\Delta \nu_1$ fit to Eq.~\ref{eq:peak-fit}. (d) Size of $G=0$ gap $\Delta\nu_2$ (measured using a cutoff of $10^{-2}\times G_0$). (inset) Predicted critical disorder $\wc(L)$ from fits of Eq.~\eqref{eq:peak-fit}.
    We use approximately $1200$ disorder realizations for $L=800$ and $300$ otherwise.
    }   
\end{figure}
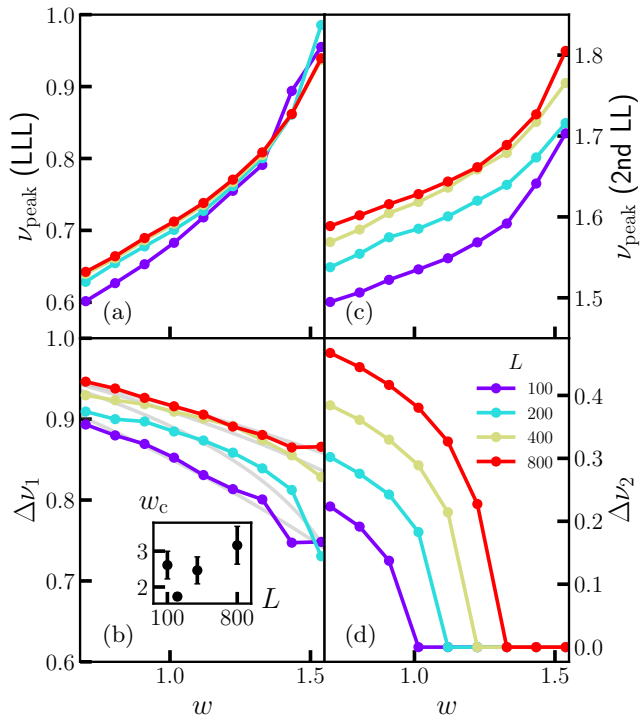

The peak position, used to calculate $\Delta \nu_1$, is calculated using a quadratic fit of a small maximal region. The zero conductance gap is measured using a conductance cutoff, which we take to be $10^{-3} \, G_0$.
In Fig.~\ref{fig:peak-positions} we show the position of the first two conductance peaks as a function of filling, as well as $\Delta \nu_1$ and $\Delta \nu_2$. 
The width of the zero conductance region $\Delta \nu_2$ goes to zero for moderate disorder due to the finite width of the Landau levels. $\Delta \nu_1$ on the other hand features a nonlinear shrinking, which we fit to a power law
\begin{equation}
\label{eq:peak-fit}
\Delta\nu_1 = (1-w/\wc (L))^{\gamma(L)}
\end{equation}
using the \textsc{curve\_fit} routine from \textsc{scipy}~\cite{scipy}. This form is designed to identify a critical disorder value $\wc(L)$ for some system size $L$ at which the extended states converge, constrained by $\Delta\nu_1(w\rightarrow 0)=1$.

\begin{figure}
    \centering
    \begin{overpic}[width=1\linewidth]{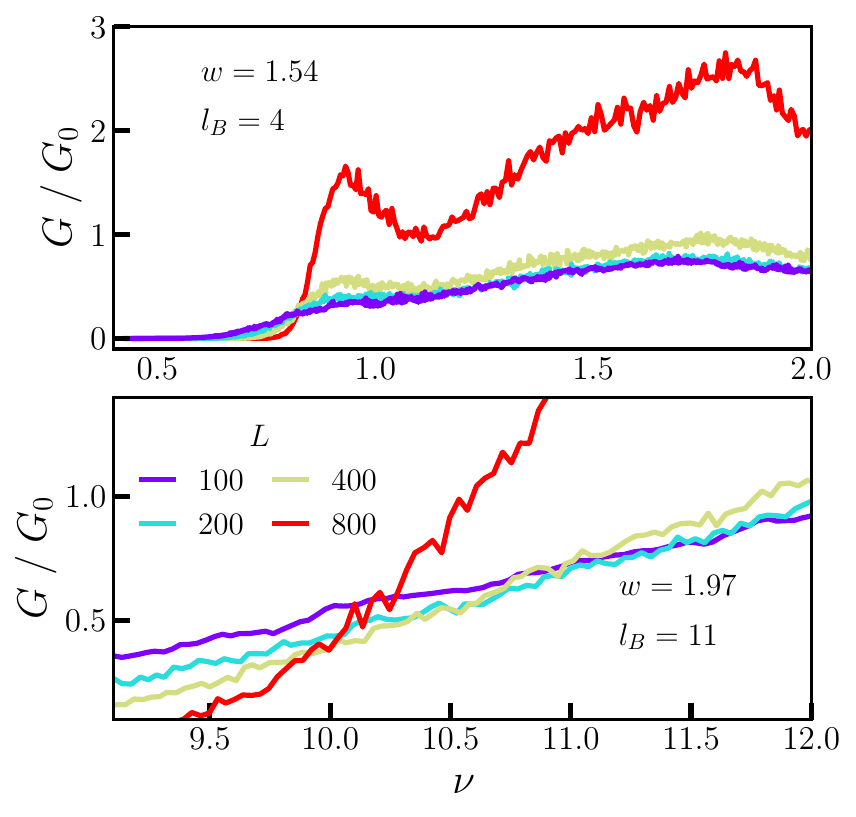}
    \put(15,88){(a)}
    \put(15,44){(b)}
    \end{overpic}
    \caption{(a) The apparent metallic phase for $L=100$ leads to a peaked profile at $L=800$. (b) The localized-to-IQHE transition point does not collapse, demonstrating the lack of a convergent transition. 
    For larger systems, we expect the crossover point to increase in $\nu$ indefinitely. 
     We use $300$ disorder realizations for $l_B=4$ and $800$ realizations for $l_B=11$.
    }
    \label{fig:localization_transition}
\end{figure}

Both $\Delta \nu_1$ and $\Delta \nu_2$ clearly show plateau shrinking with increased disorder.
Such behavior, and for that matter the existence of a critical disorder, apparently contradicts the indefinite floating picture from Fig.~\ref{fig:phase_diagram}, indicating an apparent end of floating at a finite energy. 
However, as discussed in the context of the percolation theory in Sec.~\ref{sec:percolation_model}, this discrepancy can be understood to be a finite size effect, so that we would expect the critical disorder to increase with system size. 
Of course, the percolation theory is only justified in the limit of long range disorder, which may differ from these simulations. 
Though the system size dependence of this critical disorder remains an open question, we believe that the disorder-induced slow shift of the conductance peaks to higher filling would continue indefinitely with increasing system size. We see no particular reason for the floating to end at some finite disorder strength in the thermodynamic limit, but numerically the issue remains somewhat unresolved. This scenario of continued floating with no disorder induced quantum phase transition is actually consistent with $\Delta \nu_1$ in Fig.~\ref{fig:plateau_width}(b) not vanishing for any finite disorder for all values of $L$; in fact, $\Delta \nu_1$ for our largest system ($L=800$) in Fig.~\ref{fig:floating} manifests larger values than for smaller system sizes, strongly hinting that $\Delta \nu_1$ most likely does not vanish. This is of course clearly seen also in Fig.~\ref{fig:localization_transition}(b) where the red curve corresponding to $L=800$ goes above the results for smaller sizes with increasing $\nu$, again reflecting indefinite floating in the thermodynamic limit. We note that $\Delta \nu_1$ is the more appropriate measure of the plateau width than $\Delta \nu_2$ since $\Delta \nu_2$ is strongly affected by the finite width of the conductance peaks induced by finite systems size effects.

For large disorder and high filling, Fig.~\ref{fig:floating} seems to show a metallic region of finite conductance with no obvious transport gap. However its persistence at larger systems is unclear. In Fig.~\ref{fig:localization_transition}, we focus on the length dependence of these conductance profiles at fixed disorder, demonstrating that at larger sizes a metallic profile can become peaked. We also show that there is no scale-independent crossover point between the localized regime and the IQHE regime, indicating that such a crossover occurs at the first extended state in an infinite system, and there is no quantum phase transition to zero plateau width (and thus vanishing of the zero-temperature IQHE) at some large but finite value of filling.

\subsection{Finite temperature}
\label{sec:finite-temperature}
Convolution with the Fermi-Dirac distribution allows the zero-temperature data to describe finite-temperature physics by varying the chemical potential. We calculate the filling, conductance, and density of states at a chemical potential $\mu$ and a temperature $T$ as
\begin{equation}
O(\mu,T) = - \int \frac{dE}{T} \, f'((E-\mu)/T) \, O(E)
\end{equation}
where $O$ is either the filling $\nu$, the conductance $G$ or density of states $\mathrm{DOS}$ and $f(\epsilon) = (1+e^{\epsilon})^{-1}$ is the Fermi-Dirac distribution.

\begin{figure}[t]
    \begin{overpic}[width=1\linewidth]{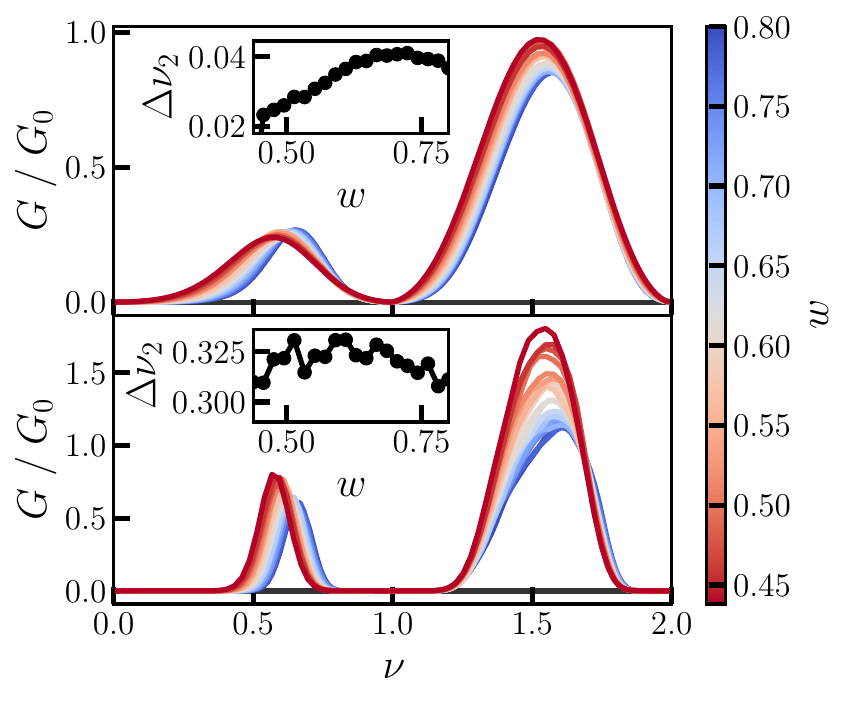}
    \put(18,55){(a)}
    \put(18,20){(b)}
    \end{overpic}
    \caption{Finite temperature plateau growth at (a) $T=0.05\hbar\omegac$ and shrinkage at (b) $T=0.01\hbar\omegac$ with increasing disorder. The $\Delta \nu_2$ gap in the insets is determined with a conductance cutoff of $10^{-3}\,G_0$, which is shown as a grey line. $L=400$.
    We use $300$ disorder realizations.}
    \label{fig:finiteT}
\end{figure}

We see in Fig.~\ref{fig:finiteT}(a) that at $T=0.05\omega_\mathrm{c}$, the plateau width, as measured by $\Delta \nu_2$, increases with increasing disorder. These results agree qualitatively with experimental results~\cite{Klitzing:1985,Furneaux:1984,Furneaux:1986,Gottwaldt:2003,mohle:1989465,Adrian:1989,Sigg:1988293,Stormer:198232,Harmand:2009} which find plateau growth and, together with the zero temperature results, demonstrate the non-monotonic relationship between plateau width and disorder strength. The reason for this increase is the shifting of the second-lowest Landau level while the point at which the conductance of the first peak vanishes remains static. However is is clear from the conductance profiles that the peak-to-peak width does not show an increasing behavior with increasing disorder since the lowest Landau level peak shifts up, unlike its tail. In Fig.~\ref{fig:finiteT}(b), a lower temperature causes the tail of the second-lowest Landau level to remain static and we see a decrease of the plateau with increasing disorder, which matches the zero-temperature case.
Thus, the dependence of the plateau width on disorder depends on the details of both temperature and the disorder strength as well as the Landau level filling; the plateau width may increase or decrease depending on the details although at $T=0$ finite disorder always shrinks the plateau. In the hypothetical limit of zero disorder, however, plateaus vanish since a spectral gap by itself is insufficient to produce IQHE; one must have a mobility gap with the chemical potential going through localized states in the gap.

\subsection{Peak scaling}
\label{sec:peak-scaling}
\begin{figure}[t]
    \includegraphics[width=1\linewidth]{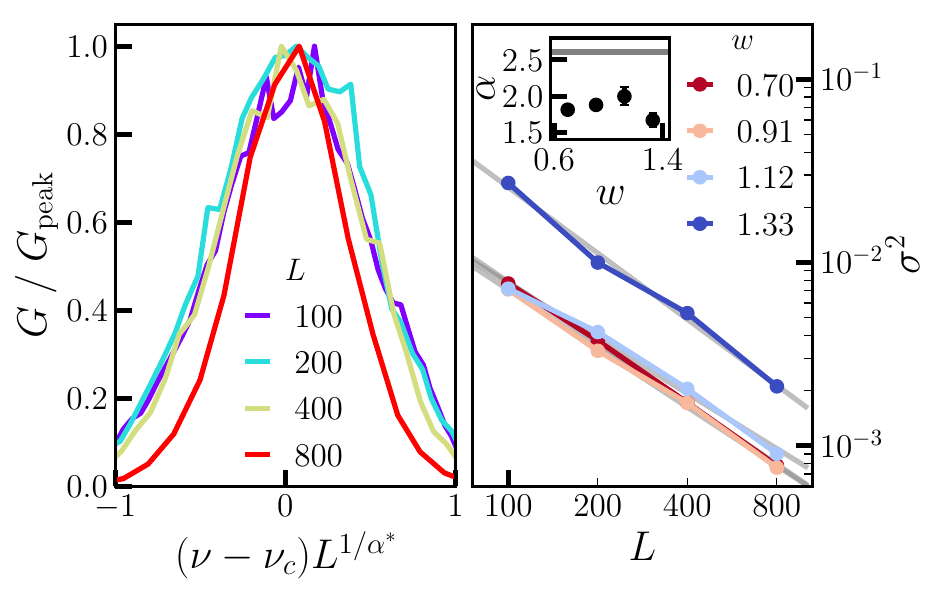}
    \caption{(Left) The scaling of the conductance peak demonstrating approximated collapse for the $w=0.7$ disorder case. We use the exponent $\alpha^*=2.609$ from Ref.~\onlinecite{dresselhaus2022}. (Right) The width $\sigma^2$ of the unscaled Gaussian calculated by fitting $\log G$ by $(\nu - \nu_c)^2$. Resulting scaling exponents are given in the inset compared with $\alpha=\alpha^*$, denoted by the horizontal line.
    We use $1200$ disorder realizations for $L=800$ and $600$ otherwise.}
    \label{fig:peak-scaling}
\end{figure}

Though the conductance peaks are expected to narrow to vanishing width in the zero-temperature thermodynamic limit, they still have considerable width for computationally feasible systems~\cite{huckestein1995}. Previous work~\cite{dresselhaus2022} estimated the length scaling of the width to have a power law relationship with a power $1/\alpha^*=1/2.609$ for an isolated Landau level. 
We use this exponent to demonstrate a scaling collapse in Fig.~\ref{fig:peak-scaling}. 
We then calculate specific scaling exponents for each disorder value and compare them to the $1/2.609$, finding rough agreement ($\approx 1/2$). 
It is worth noting that the mixing of Landau levels, which is included in our exact calculation, should modify the length scaling and thus we do not expect perfect agreement with the exponent extracted for an isolated single Landau level in Ref.~\onlinecite{dresselhaus2022}.
This discrepancy could also be due to higher-order terms in studied in Ref.~\onlinecite{huckestein1995}. Importantly, we find a power law relationship which suggests that the peaks do in fact narrow with increasing system size, though at a slow rate.
The near quantitative agreement of this result with those obtained from the network model~\cite{dresselhaus2022} 
justifies the use of the percolation limit to extrapolate these results to the thermodynamic limit discussed in the following sections since the network model is essentially the lattice version of the percolation model.

\section{Percolation model}\label{sec:percolation_model}
The exact results presented in Sec.~\ref{sec:tight-binding} bring out all the essential features of disorder and temperature effects on IQHE, but suffer from finite size effects since such calculations obviously cannot be extended to infinite system size.  
In this section, we use an approximate analytical theory based on the percolation model~\cite{Kazarinov:1982,Trugman:1983,Prange:1982,Joynt:1984}, which is closely related to the so-called network model~\cite{KRAMER2005211}, which are both extensively used in theoretical works on IQHE.  The model, to be described below, assumes slowly varying potential with the disorder correlation length being much larger than the magnetic length.  The model thus becomes increasingly more accurate at high magnetic fields and/or for long range potential  arising from Coulomb disorder.  The percolation model enables obtaining approximate results in the thermodynamic limit.  
The models, approximations, limitations,  and the theoretical techniques of Sec.~\ref{sec:tight-binding} and \ref{sec:percolation_model} are thus complementary, but they lead to the same qualitative conclusions about the effects of disorder and temperature on IQHE.

In this section, we study how temperature and disorder
affect the percolation model of IQHE~\cite{Kazarinov:1982,Trugman:1983,Prange:1982,Joynt:1984,KRAMER2005211}.
The main results of this section are summarized as follows.
At zero temperature, the plateau width decreases as disorder increases due to the floating effect [cf.\ Figs.~\ref{fig:plateau_width} (a-b)], consistent with the $T=0$ results from tight-binding simulations in Sec.~\ref{sec:tight-binding}.
At a finite temperature $T$ where $T<\hbar \omega_c$, the plateau width shows a non-monotonic dependence on disorder.
The plateau width initially increases with disorder for small disorder strengths due to weaker thermal activation.
However, at higher disorder, the plateau width narrows again due to floating [cf.\ Fig.~\ref{fig:plateau_width_schematic} and \ref{fig:plateau_width} (c-d)].
These results are in agreement with the tight-binding simulations shown in Fig.~\ref{fig:finiteT}, where for a small (large) effective disorder $w/T$, the lowest plateau width increases (decreases) as $w$ increases.
A key qualitative finding is that for temperature larger than disorder, the IQHE is essentially suppressed, indicating that IQHE would be more challenging to observe in high-quality low-disorder samples because one must go to very low temperatures to observe IQHE.  The fact that IQHE is generally more prominent in higher disorder samples has been known since the early days~\cite{Stormer:198232} as discussed below.

The effect of disorder on plateau widths has been investigated in experiments. 
For example, as shown in Fig.~4 of Ref.~\cite{Klitzing:1985}, the plateau is better developed for a GaAs sample with lower mobility $10^5$ cm$^2$/Vs than the other sample with higher mobility $10^6$ cm$^2$/Vs, despite both samples having the same carrier concentration and being measured at the same temperature.
A similar observation is reported in Fig.~7 of Ref.~\cite{Harmand:2009}, where a GaAs sample doped by a $\delta$-layer of Be with lower mobility $0.36\times10^5$ cm$^2$/Vs shows a wider plateau than higher mobility samples with mobilities of $0.8\times10^5$ cm$^2$/Vs and $5\times10^5$ cm$^2$/Vs.
All samples in Ref.~\cite{Harmand:2009} have a similar carrier concentration $\approx 2.5 \times 10^{11}$ cm$^{-2}$ and the same structure parameters, with the only difference being the concentration of impurities in the $\delta$ layer.
In addition, a comparable trend with a wider plateau for lower transport mobility or effectively larger disorder is observed in modulation-doped GaAs quantum wells~\cite{Gottwaldt:2003}, irradiated GaAs quantum wells~\cite{mohle:1989465,Adrian:1989,Sigg:1988293}, Si MOSFET doped with driftable Na$^+$ ions as shown in Fig. 6 of Ref.~\onlinecite{Furneaux:1986} or Fig. 3 of Ref.~\onlinecite{Furneaux:1984}.
Other experimental data which changes the mobility by tuning electron concentration can be found in Ref.~\onlinecite{Stormer:198232}.
For samples in Ref.~\onlinecite{Furneaux:1986} with mobilities higher than $10^4$ cm$^2$/Vs, the plateau width increases as the mobility decreases. 
However, for lower mobility samples in Ref.~\onlinecite{Furneaux:1986} ($<10^4$ cm$^2$/Vs), the opposite trend is observed, with the plateau width decreasing as the mobility decreases.
Rather dramatically, the extreme high-mobility ($\sim 3\times10^7$ cm$^2$/Vs) GaAs samples of Refs.~\onlinecite{Myers:2021,Csathy_ultrahigh:2024,Myers:2024} manifest decent IQHE plateaus only at extremely low temperatures ($\sim 10$ mK) with the plateaus basically disappearing for $T=100$ mK.  
On the other hand, for the low-mobility ($\sim 5 \times 10^4$ cm$^2$/Vs) GaAs samples reported in Refs.~\onlinecite{Razeghi:1985,Tsui:1988,Tsui:1992,Klitzing:1991}, IQHE plateaus persist up to $\sim 10$ K.
Our theory provides an explanation for this striking observation.
These observations suggest that samples can be classified into two categories: in high-mobility samples with relatively small disorder, the plateau width increases as mobility decreases, while in low-mobility samples, the plateau width decreases as mobility decreases.
This experimental observation qualitatively agrees with our result shown in Fig.~\ref{fig:plateau_width_schematic}.

In Sec.~\ref{subsec:classical_localization}, we describe the percolation model of the IQHE in the presence of a long-range disorder potential. 
In this model, all states are classically localized when the magnetic field is sufficiently strong, except for a delocalized extended state at the center of each Landau level. 
The percolation of this extended state is ensured by the particle-hole symmetry of the disorder potential.
Typically, the percolation model becomes very accurate for modulation-doped samples where the distant charged impurities create a slowly varying long-range disorder potential.  
We establish the accuracy of the percolation model for such a long-range disorder by calculating the drift of the guiding center of the cyclotron orbit, establishing its percolating character.
In Sec.~\ref{subsec:plateau-width-finite-T}, we compute the plateau width as a function of disorder and temperature using the percolation model.
At finite temperatures, thermal activation delocalizes an energy band $\sim T$ around the center of the LL, so that the plateau width is given by (heuristically)
\begin{align}\label{eq:analytical_plateau_width}
    \Delta \nu (T) \approx \Delta \nu(T=0)- \frac{T}{\min\{\Gamma,\hbar \omega_c/2\}},
\end{align}
where $\Gamma$ is the LL broadening in the density of states (DOS) which characterizes the strength of the long-range disorder. 
(As we explain later in this section, the broadening parameter $\Gamma$ in general is different from the bare disorder potential $w$ defined in Section~\ref{sec:tight-binding}, since the bare disorder is screened by the carriers and averaged over fast cyclotron motion).
We perform numerical calculations of $\sigma_{xy}(\nu,T,\Gamma)$, obtaining the related plateau width $\Delta \nu$ which verifies Eq.~\eqref{eq:analytical_plateau_width}.
In Sec.~\ref{subsec:VRH}, we discuss the competition between thermal activation and hopping that leads to variable range hopping (VRH) at sufficiently low temperatures.

\subsection{Magnetic field induced classical localization and percolation in long-range disorder potential} \label{subsec:classical_localization}
For a long-range disorder potential $U(\vb{r})$, characterized by the root mean square fluctuation $\delta U$ and a correlation length $d$, classical localization occurs when the cyclotron radius $R_c=v_F/\omega_c$ is sufficiently small, where $v_F$ is the Fermi velocity of the electron. 
(For a short-range disorder, $\delta U$ is represented by $w$ in Section~\ref{sec:tight-binding}).
In this scenario, the electron orbit drifts along certain equipotential lines of the long-range disorder potential, becoming confined within these contours.
We first compute the scattering rate using the Born approximation, which is valid for $E_F \gg \delta U$.
For example, this long-range potential can be generated by random remote charged impurities in a $\delta$-layer at a distance $z=d$ away from the 2DEG, such that
\begin{align}
    U(\vb{r}) = \sum_{i=1}^{N_i} U_1(\vb{r}- \vb{R}_i, d),
\end{align}
where $\vb{R}_i$ is the $i$-th 2D coordinate of an impurity charge, $N_i$ is the total number of impurities corresponding to some finite 2D concentration $n_i = N_i/A$, $A$ is the total area of the 2D system, and $U_1(\vb{r},d)$ is the single-impurity potential.
For example, the unscreened Coulomb potential reads
\begin{align}
    U_1(\vb{r},d) = \frac{e^2}{\kappa \sqrt{\abs{\vb{r}}^2 + d^2}} = \frac{1}{A} \sum_{\vb{q}} e^{i\vb{q} \cdot \vb{r}} U_{\vb{q}}.
\end{align}
The Fourier component of the Coulomb potential screened by the 2DEG is given by
\begin{align}\label{eq:uq}
    U_{\vb{q}} = \frac{2\pi e^2}{\kappa (q + q_{\mathrm{TF}})} e^{-qd},
\end{align}
where $q_{\mathrm{TF}} = g_s g_v/a_B$ is the Thomas-Fermi wavevector with $g_s$ ($g_v$) the spin (valley) degeneracy.
(For the rest of this section we assume $g_s=g_v=1$.)
$a_B = \kappa \hbar/ m e^2$ is the effective Bohr radius with $\kappa$ the effective background lattice dielectric constant. 
Specifically, for unscreened Coulomb potential $q_{\mathrm{TF}} = 0$.
The drift velocity is given by
\begin{align}\label{eq:v_d}
    v_d \approx \frac{\grad U}{m \omega_c}\sim \frac{\delta U}{m \omega_c d},
\end{align}
where $\delta U$ is the root-mean-square fluctuation of the random potential $U(\vb{r})$ defined as
\begin{align}\label{eq:bare_disorder_potential_square}
    \ev{(\delta U)^2}_d \equiv \ev{U(\vb{r})^2}_d - \ev{U(\vb{r})}_d^2,
\end{align}
and the disorder average is performed through the impurity coordinates
\begin{align}
    \ev{O(\{\vb{R}_i\})}_d = \prod_{\{\vb{R}_i\}} \int d^2 R_i O(\{\vb{R}_i\}).
\end{align}
As a result, after disorder average we have
\begin{align}\label{eq:w2}
    (\delta U)^2 = n_i \int d^2 r U_1(\vb{r})^2 = n_i \frac{1}{A} \sum_{\vb{q}} \abs{U_{\vb{q}}}^2.
\end{align}
Substituting Eq.~\eqref{eq:uq} into Eq.~\eqref{eq:w2}, we obtain
\begin{align}
    (\delta U)^2 &= 2\pi n_i \qty(\frac{e^2}{\kappa})^2 [-1 - (1 + 2 q_{\mathrm{TF}} d) \mathrm{Ei}(-2q_{\mathrm{TF}} d) e^{2q_{\mathrm{TF}} d}], \nonumber \\
    &= 2\pi n_i \qty(\frac{e^2}{\kappa})^2
    \begin{cases}
        -1-\gamma - \ln (2q_{\mathrm{TF}} d), & q_{\mathrm{TF}} d \ll 1, \\
        [4 (q_{\mathrm{TF}} d)^2]^{-1}, & q_{\mathrm{TF}} d \gg 1,
    \end{cases}\label{eq:remote_potential_square}
\end{align}
where $\mathrm{Ei}(x)$ is the exponential integral function.
We are interested in the long-range potential such that $q_{\mathrm{TF}} d \gg 1$ so that the percolation condition of a sufficiently slowly varying potential is satisfied.
According to Fermi's golden rule, the transport scattering rate should be proportional to $(\delta U)^2$ through
\begin{align}
    \frac{1}{\tau} &= \frac{2\pi}{\hbar} \sum_{\vb{k}_f} (1-\cos\theta) \delta(\varepsilon_{\vb{k}_f} - E_F) \ev{\abs{\ev{\vb{k}_{i}| \delta U| \vb{k}_{f} }}^2}_d, \nonumber \\
    &=\frac{2\pi}{\hbar} \frac{m}{2\pi \hbar^2} n_i \int_0^{2\pi} \frac{d\theta}{2\pi} \abs{U_{\vb{q}}}^2 (1-\cos\theta), \nonumber \\
    &\approx \qty(\frac{2}{g_s g_v})^2 \frac{\pi \hbar n_i}{m (2k_F d)^3} = \frac{m (\delta U)^2}{\hbar^3 k_F^3 d},
\end{align}
where in the first step, $\vb{k}_i$ ($\vb{k}_f$) is the momentum of the initial (final) state of the electron, with the (elastic) scattering angle $\theta$ and the momentum transfer $q = 2k_F \sin (\theta/2)$.
The scattering potential with disorder average is defined as
\begin{align}
    \ev{\abs{\ev{\vb{k}_{i}| \delta U| \vb{k}_{f} }}^2}_d &\equiv \ev{\abs{\ev{\vb{k}_{i}| U| \vb{k}_{f} }}^2}_d - \abs{\ev{\ev{\vb{k}_{i}| U| \vb{k}_{f} }}_d}^2, \nonumber \\
    &= \frac{n_i}{A} \abs{U_{\vb{q}}}^2.
\end{align}
The quantum (single-particle) scattering time in the limits of $q_{\mathrm{TF}} d, k_F d \gg 1$ is given by
\begin{align}\label{eq:tau_q}
    \frac{1}{\tau_q} &\approx \qty(\frac{2}{g_s g_v})^2 \frac{\pi \hbar n_i}{2 m k_F d} = \frac{4 m (\delta U)^2 d}{\hbar^3 k_F},
\end{align}
If $\omega_c \tau \gg 1$ and $R_c \gg d$, an electron moves fast so that it completes many revolutions between adjacent scattering events, whose trajectory covers many uncorrelated potential regions of size $\sim d$.
The drift velocity for each uncorrelated potential regime fluctuates in both direction and magnitude, so the guiding center performs a random walk of a time step $\delta t \sim d/v_F \ll \omega_c^{-1} \ll \tau$ and a random velocity with a typical magnitude given by $v_d$ in Eq.~\eqref{eq:v_d}.
Each step of the random walk displaces the guiding center by $\delta r \sim v_d \delta t = v_d (d/v_F)$.
After some time $t\gg \delta t$, the total displacement due to random walk should be $\delta r \sqrt{t/\delta t}$.
Specifically, the displacement of the guiding center after one cyclotron period aligns with the result derived in Refs.~\cite{Baskin:1978,Laikhtman:1994,Fogler:1997}
\begin{align}%
    \delta r_c \approx \frac{v_d}{v_F} d \sqrt{\frac{R_c}{d}} = \frac{v_d}{v_F} R_c \sqrt{\frac{d}{R_c}} = \frac{R_c}{\sqrt{\omega_c \tau}}.
\end{align}
In other words, if $R_c \gg d$, the fast revolution of an electron self-averages the fluctuating disorder potential and reduces $\delta U$ by a factor of $\sqrt{d/R_c}$ in a cyclotron period.
The result of $(\delta U_{\mathrm{eff}})^2 =(\delta U)^2 (d/R_c)$ for $R_c \gg d$ is the same as Eq.~(57) in Ref.~\cite{Raikh:1993}.
On the other hand, if $R_c < d$, then $\delta U_{\mathrm{eff}} \approx \delta U$ remains the same after the self-averaging.

When $\delta r_c<d$, diffusion is insufficient to shift an electron between separate regions of size $d$ in the disorder potential. Consequently, the electron becomes confined along an equipotential line within a domain of size $d$. The diffusion coefficient is exponentially small, given by $D_B \sim \omega_c d^2 e^{-B/B_c}$, where $B_c$ is the magnetic field for which $\delta r_c = d$~\cite{Fogler:1997}
\begin{align}
    B_c = \frac{mc}{e \hbar} \qty(\frac{\hbar^2}{m d^2})^{1/2} \frac{(\delta U)^{2/3}}{E_F^{1/6}}.
\end{align}
In other words, the diffusion is exponentially suppressed if $R_c$ (or $E_F$) is sufficiently small such that $E_F < E_{Fc}$ with
\begin{align}\label{eq:E_Fc}
    E_{Fc} = \qty(\frac{\hbar^2}{m d^2})^{3} \frac{(\delta U)^{4}}{(\hbar \omega_c)^6}.
\end{align}
This phenomenon, known as classical localization in a smooth random potential under a magnetic field, may explain the quantized Hall conductance plateau, which requires a significant reduction in longitudinal conductivity~\cite{Fogler:1997}.
At the center of each LL, an equal-energy contour always percolates through the entire system. 
(This is in fact the saddle point of the disorder landscape characterizing an extended state percolating through the whole disorder network.)
This percolation is guaranteed by the particle-hole symmetry of the disorder potential, where $\delta U(\vb{r})$ is statistically equivalent to $-\delta U(\vb{r})$. 
In other words, the probability distributions of ``valley'' and ``mountains'' in the long-range disorder potential are identical (i.e., a saddle point). 
As a result, for sufficiently strong magnetic fields ($B > B_c$ or $E_F < E_{Fc}$), all states are localized, except for the extended states that percolate at the center of each LL.
The percolation scenario thus naturally produces IQHE at strong enough magnetic field  (or equivalently, small enough Fermi energy) values because of the existence of localized states everywhere except at the LL centers where extended states exist.

On the other hand, for $B < B_c$ or $E_F > E_{Fc}$, the diffusion of the electron orbit guiding center becomes strong enough for electrons to traverse the entire disorder landscape, and the system transitions into a diffusive metal where both $\sigma_{xx}$ and $\sigma_{xy}$ follow the classical Drude formula.
In this `classical' Drude low-field or high-density regime, the standard pre-1980 textbook magnetotransport prevails with NO IQHE and/or Shubnikov-de Haas (SdH) oscillations in the longitudinal magnetoresistance with the Hall conductance given simply by the classical Hall formula proportional to $n/B$,  where $n$ is the 2D electron density.  Experimentally, this regime is routinely observed experimentally in the weak field regime, where the high field manetotransport oscillations cross over to the semiclassical monotonic results with the Landau level structure suppressed in transport.
We assume that the system size is smaller than the exponentially long Anderson localization length so that the 2D system is apparently metallic if disorder is not very strong.
Low-field experimental phenomenology is well described by the Drude response of an effective metallic Fermi liquid.
For the low-field regime $B<B_c$, two localization mechanisms -- quantum Anderson localization and semiclassical percolation localization -- may compete in the presence of sufficiently large disorder. 
Anderson localization arises from destructive quantum interference between multiple-scattering paths induced by disorder~\cite{anderson1958,ioffe1960}, while percolation localization results from the fragmentation of the 2DEG into inhomogeneous electron puddles separated by long-range Coulomb disorder potential barriers~\cite{Thouless:1971,Kirkpatrick:1973,Shklovskii:1975,shklovskii:2024}.
This competition frequently occurs in zero-field experimental systems, making it challenging to identify the dominant mechanism underlying the metal-insulator transition~\cite{DasSarma:2005579,Tracy:2009,Manfra:2007,Qiuzi:2013,Tracy:2014,DasSarma:2005,Ahn:2022,Poduval:2023,Huang:2021,huang2023understanding,Huang_TMD:2024}.
While the interplay between Anderson and percolation localization in the low-field regime is an important topic, it is beyond the scope of this work, where our interest is strictly on disorder effects on IQHE, which, for the percolation model, is explicitly the strong field regime with extended states percolating through the system only at the Landau level centers. 
Therefore, in this section, we focus on the high-field and low-density regime $B>B_c$ and $E_F < E_{Fc}$, where the IQHE percolation model, involving magnetic-field-induced classical localization between plateau transitions, provides a justified framework for analysis.

In the presence of disorder potential $U(\vb{r})$, using self-consistent Born approximation (SCBA), the singular delta-function DOS of LLs gets broadened~\cite{Ando:1974}
\begin{align}\label{eq:LL_DOS_semicircle}
    g(\varepsilon) = \frac{1}{2\pi l_B^2} \sum_N \frac{2}{\pi \Gamma}\sqrt{1-\frac{(\varepsilon - E_N)^2}{\Gamma^2}} \Theta(\Gamma - \abs{\varepsilon - E_N}),
\end{align}
where $E_N = N\hbar \omega_c$, $N=1,2,3,\dots$, is the energy of the $N$-th LL (we shift the Hamiltonian relative to the zero point energy $\hbar \omega_c/2$).
The semicircle shape of DOS with a hard gap and unphysical infinite slope at the sharp edges is an artifact of SCBA which only includes leading-order disorder scattering, and the DOS, especially in the case of long-range disorder potential, is better described by a Gaussian shape with the well-known disorder-induced band tailing effect~\cite{Raikh:1993,Laikhtman_Altshuler:1994}
\begin{align}\label{eq:LL_DOS_gaussian}
    g(\varepsilon) = \frac{1}{2\pi l_B^2} \sum_N \frac{1}{\sqrt{2\pi \Gamma^2}} e^{-\frac{(\varepsilon-E_N)^2}{2\Gamma^2}}.
\end{align}
The level broadening $\Gamma$ in the LL DOS is given by
\begin{align}
    \Gamma^2 &= \frac{n_i}{2\pi^2 R_c} \int_0^{\infty} dq \abs{U_{\vb{q}}}^2, \\
    &= 2n_i \qty(\frac{e^2}{\kappa})^2 \frac{d}{R_c} \qty[\frac{1}{q_{\mathrm{TF}} d} + 2 e^{2 q_{\mathrm{TF}} d} \mathrm{Ei}(- 2 q_{\mathrm{TF}} d)], \\
    &= 2n_i \qty(\frac{e^2}{\kappa})^2 \frac{d}{R_c}
    \begin{cases}
        (q_{\mathrm{TF}} d)^{-1}, &\qif q_{\mathrm{TF}} d\ll 1, \\
        [2(q_{\mathrm{TF}} d)^2]^{-1}, &\qif q_{\mathrm{TF}}  d\gg 1. \label{eq:gamma_square_long_range_SCBA}
    \end{cases}
\end{align}
where $U_{\vb{q}}$ is given by Eq.~\eqref{eq:uq}.
For long-range disorder in the limit of $q_{\mathrm{TF}}  d\gg 1$, we can compare Eqs.~\eqref{eq:gamma_square_long_range_SCBA} and \eqref{eq:remote_potential_square} and obtain
\begin{align}
    \Gamma^2 = \frac{2 }{\pi} \frac{d}{R_c} (\delta U)^2.
\end{align}
where $(\delta U)^2$ for remote impurities of a $\delta$-layer at a distance $z=d$ away from the 2DEG is given by Eqs.~\eqref{eq:bare_disorder_potential_square} and \eqref{eq:remote_potential_square}.
This result can be understood physically as a result of self-average of disorder potential within the length scale of the cyclotron radius $R_c \gg d$.
Comparing Eqs.~\eqref{eq:gamma_square_long_range_SCBA} and \eqref{eq:tau_q}, we obtain the expression as derived in SCBA of Refs.~\cite{Ando:1974,Raikh:1993,Laikhtman_Altshuler:1994} up to a numerical factor
\begin{align}
    \Gamma^2 = \frac{1}{2\pi} \hbar \omega_c  \frac{\hbar}{\tau_q}.
\end{align}
By contrast, if $R_c \ll d$, then the self-average in $R_c$ gives the same bare potential so that $\Gamma^2 \sim (\delta U)^2 = (\hbar^2 k_F/4md) (\hbar /\tau_q)$ which is independent of the magnetic field.
We note that the LL broadening $\Gamma$ increases as the disorder increases, being proportional to the imaginary part of the electron self-energy ($\hbar/\tau_q$) due to disorder.
Therefore, in the rest of this section, we use $\Gamma$ as a parameter to characterize the strength of the disorder.
As an aside, we mention that while the broadening parameter $\Gamma$ for short-range disorder is basically given by the transport scattering rate $1/\tau$ as manifested in sample mobility, $\Gamma_q=\hbar/\tau_q$ may far exceed the mobility scattering rate $\hbar/\tau$ for long-range disorder, since $\tau\gg \tau_q$ for long-range disorder by virtue of vertex corrections from forward scattering whereas for short-range disorder $\tau\sim \tau_q$ due to the s-wave nature of short-range scattering~\cite{Stern:1985,Dmitriev:2012}.  
For us here in the current work, however, $\Gamma$ is a tunable disorder parameter denoting which is varied to ascertain its effect on the IQHE plateau.

\subsection{Quantum Hall plateau width as a function of disorder at finite temperatures}
\label{subsec:plateau-width-finite-T}
At low temperatures, only states near the center of the LL around $E_N=N \hbar \omega_c$ are delocalized so that $\sigma_{xx}\neq0$ and $\sigma_{xy}$ is not quantized around the LL center. 
In the following, we define the effective plateau width that corresponds to $\sigma_{xy}=N e^2/h$ as 
\begin{align}\label{eq:delta_nu_N}
    \Delta \nu (N) = &\nu(N+1) - \nu(N) \nonumber\\
    &- [\Delta\nu_p(N+1) + \Delta\nu_p(N)]/2.
\end{align}
where the width of the $\sigma_{xx}$ peak in the filling factor is denoted as $\Delta \nu_p$, and $\nu(N)$ is the $\sigma_{xx}$ peak position.
We analytically calculate $\Delta \nu_p (T)$ and $\nu(N)$ at a finite temperature for the percolation model described in Section~\ref{subsec:classical_localization}, assuming that at $T=0$ only a single state in the center of LL is delocalized, while all other states are strongly localized with very small localization length such that at finite $T$ delocalization is entirely induced by thermal activation, and hopping effects can be ignored.
(We consider hopping effects later in Sec.~\ref{subsec:VRH}.)
The expected $\Delta \nu(N,\Gamma,T)$ as a function of $\Gamma$ at a finite $T\ll \hbar \omega_c$ is plotted schematically in Fig.~\ref{fig:plateau_width_schematic}.

To summarize the qualitative aspects of the finite temperature results, the plateau width exhibits a non-monotonic dependence on disorder.
Due to thermal activation, $\Delta \nu\approx 1-(T/\Gamma)$ increases as $\Gamma$ increases for weak disorder $\Gamma < \hbar \omega_c$.
Specifically, a plateau disappears completely if $\Gamma < T$.
For the intermediate disorder $\hbar \omega_c < \Gamma< N \hbar \omega_c$, $\Delta \nu \approx 1-(T/\hbar \omega_c)$ saturates to a constant independent of the disorder.
For large disorder $\Gamma > N \hbar \omega_c$, $\Delta \nu$ decreases as $\Gamma$ increases because of the accumulation of localized states below the lowest LL.
In other words, as disorder increases, LLs float to higher fillings, and the spacing between LLs shrinks.
This floating result in the percolation model is consistent with the tight-binding simulation discussed in Section~\ref{sec:tight-binding}.
We also numerically compute $\sigma_{xy}(\nu)$ for various $T$ and $\Gamma$ that verifies the above results. 
In Section~\ref{subsec:VRH}, we discuss the competition between thermal activation and hopping that leads to variable range hopping (VRH).

\begin{figure}
    \centering
    \includegraphics[width=0.8\linewidth]{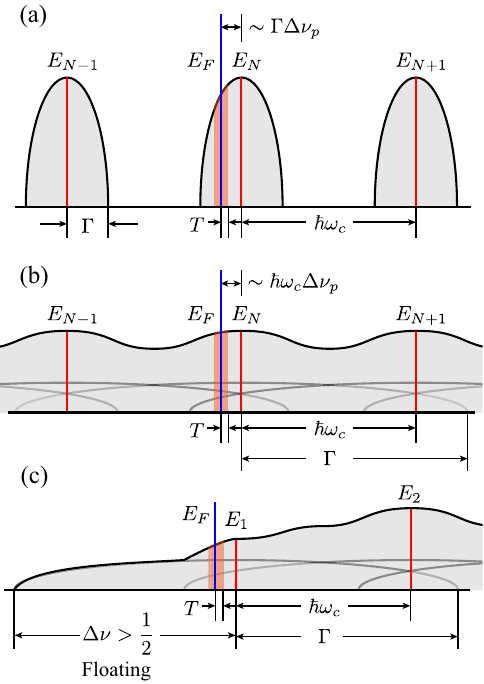}   
    \caption{Schematics of density of states (DOS) as a function of energy to illustrate different energy scales $\Gamma$, $\hbar \omega_c$, $\abs{E_F - E_N}$, and $T$ for broadened Landau levels (LLs). The red vertical lines represent the delocalized states at the center of LLs. The blue vertical line represents the postion of Fermi level $E_F$. The orange shaded regime near $E_F$ represents the thermal excitations with probability of order 1, while outside this regime the probability is exponentially suppressed by the energy difference. (a) The half-width of LL broadening $\Gamma$ is smaller than $\hbar \omega_c$ so LLs do not overlap, and $\abs{E_F - E_N}\sim \Gamma \Delta \nu_p$, where $\Delta \nu_p$ is the filling factor that deviates from the center of a LL. (b) $\Gamma > \hbar \omega_c$. The black curve represents the total DOS while the light gray curves represent the DOS corresponding to individual LLs which overlap with each other. In this case, the total DOS is roughly a constant with small sinusoidal modulation, and $\abs{E_F - E_N}\sim \hbar \omega_c \Delta \nu_p$. (c) An example of floating where the center of the lowest LL shifts to a filling factor larger than $1/2$ when disorder is large $\Gamma >\hbar \omega_c$. In general, floating of the $N$-th LL occurs when $\Gamma > N \hbar \omega_c$.}
    \label{fig:LL_DOS}
\end{figure}

\begin{figure}
    \centering
    \includegraphics[width=0.9\linewidth]{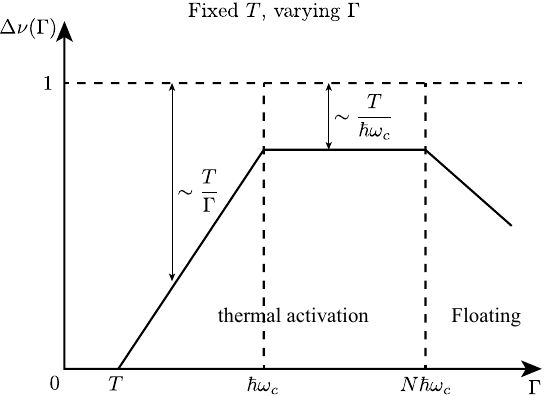}
    \caption{Schematic plot of the plateau width $\Delta \nu$ [cf.\ Eq.~\eqref{eq:delta_nu_N}] corresponding to the $N$-th QH plateau $\sigma_{xy}=N e^2/h$ as a function of disorder $\Gamma$ at a fixed temperature $T$. Note that this figure is consistent with the plateau width always decreasing with increasing disorder at $T=0$-- increasing width with increasing disorder necessitates having a finite $T$.  For sufficiently large $T$, the plateaus are always totally suppressed.}
    \label{fig:plateau_width_schematic}
\end{figure}

We compute the probability that an electron at the Fermi level is thermally activated above the mobility gap, i.e., across the center of the LL.
Such thermally excited electrons become `mobile', and do not contribute to IQHE, thus inducing a thermal suppression of the plateau if all other parameters are held fixed.
Suppose that the Fermi level $E_F$ is somewhere between $E_{N-1}$ and $E_N$, the electron density thermally excited above $E_N$ is given by
\begin{align}\label{eq:activated_nT}
    \Delta n_e (T,E_F,\Gamma) = \int_{E_N}^{\infty} d\varepsilon g(\varepsilon) f\qty(\frac{\varepsilon-E_F}{T}),
\end{align}
where $f(x) = (e^{x}+1)^{-1}$ is the Fermi-Dirac distribution function.
For simplicity, we first consider the case where the LLs do not overlap $\Gamma < \hbar \omega_c $, such that only the $N$-th LL contributes to the integral. The case of strongly overlapping LLs where $\Gamma \gg \hbar \omega_c$ will be discussed later.
For non-overlapping LLs, using the DOS Eq.~\eqref{eq:LL_DOS_semicircle}, Eq.~\eqref{eq:activated_nT} can be rewritten as
\begin{align}
    \Delta n_e (T,E_F,\Gamma) &= \int_{E_N}^{E_N+\Gamma} d\varepsilon g(\varepsilon-E_N) f\qty(\frac{\varepsilon-E_F}{T}), \\
    &= \frac{1}{2\pi l_B^2} P\qty(\frac{\varepsilon-E_F}{T},\frac{T}{\Gamma}),
\end{align}
where
\begin{align}
    P(z,a) = \frac{2}{\pi} \int_0^1 dx \frac{(1-x^2)^{1/2}}{z e^{x/a} + 1},
\end{align}
represents the probability for an electron to be thermally excited across the center of LL. 
The asymptotic limits of $P(z,a)$ can be evaluated analytically.
If $z = (\varepsilon-\varepsilon_F)/T \gg 1$, then 
\begin{align}
    P(z\gg1,a) = 
    \begin{cases}
        \frac{1}{2(e^{z} + 1)}, &\qif a=\frac{T}{\Gamma} \gg 1, \\
        \frac{2a}{\pi} e^{-z}, &\qif a=\frac{T}{\Gamma} \ll 1,
    \end{cases}
\end{align}
where the exponential factor $e^{-z}$ characterizes the thermal activation.
On the other hand, if $z = (\varepsilon-\varepsilon_F)/T \ll 1$, then 
\begin{align}
    P(z\ll1,a) = 
    \begin{cases}
        \frac{1}{4}, &\qif a=\frac{T}{\Gamma} \gg 1, \\
        \frac{2a}{\pi} \ln(2), &\qif a=\frac{T}{\Gamma} \ll 1.
    \end{cases}
\end{align}
Physically, this means that for $\abs{E_F - E_N} \simeq \Gamma \Delta \nu_p \le T$ or 
\begin{align}\label{eq:delta_nu_activation}
    \Delta \nu_p = \frac{T}{\Gamma}, \qif T< \Gamma <\hbar \omega_c,
\end{align}
electrons can be thermally activated to the center of the LL, and thus be delocalized.
The corresponding plateau width $\Delta \nu \approx 1-\Delta \nu_p$ then increases as disorder increases.
The physical picture for thermal activation is illustrated in Fig.~\ref{fig:LL_DOS}(a), where a range of filling factors $\Delta \nu_p=T/\Gamma$ is delocalized if $\abs{E_F - E_N} <T$.

\begin{figure*}
    \centering
    \includegraphics[width=0.75\linewidth]{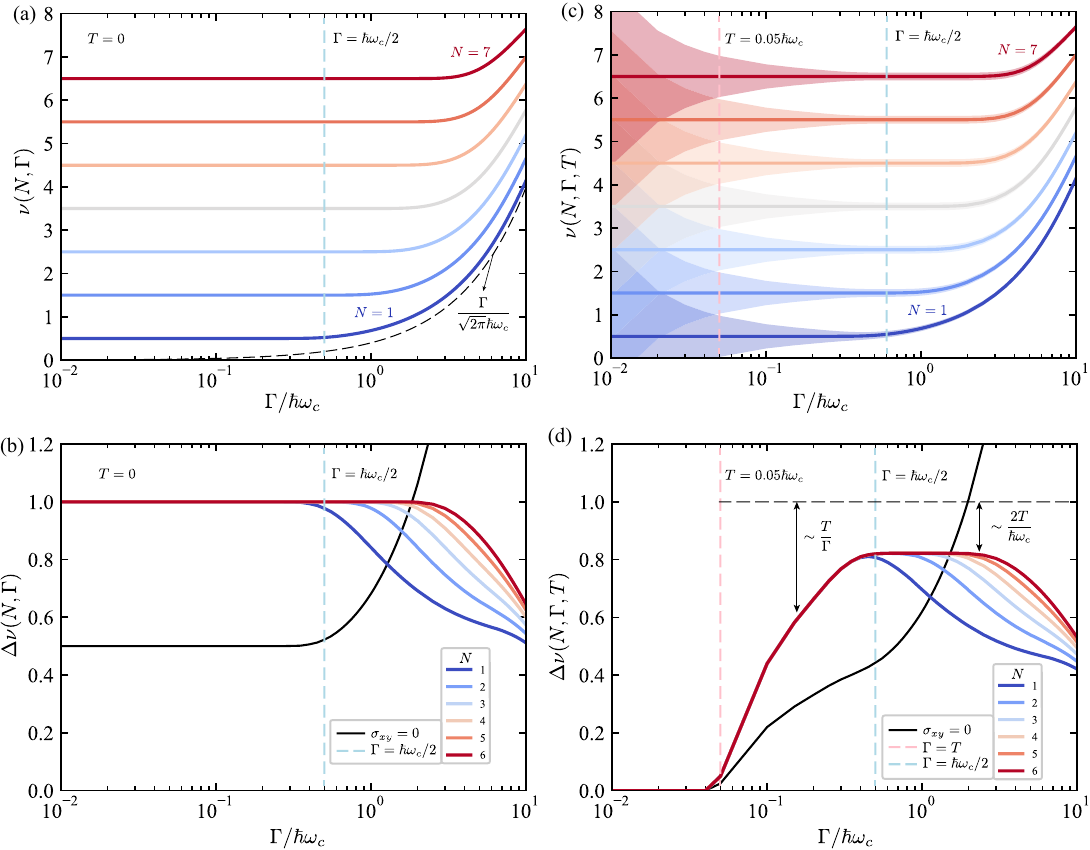}
    \caption{$T=0$ result of (a) the plateau transition position $\nu(N, \Gamma)$ that corresponds to the center of $N$-th Landau level [cf.\ Eq.~\eqref{eq:nu_N_T0}], and (b) the quantum Hall plateau width $\Delta\nu(E_N, \Gamma) = \nu(N+1, \Gamma) - \nu(N, \Gamma)$ [cf.\ Eq.~\eqref{eq:plateau_width_T0}] where $\sigma_{xy}=N e^2/h$. The black dashed curve in (a) represents $\Gamma/(\sqrt{2\pi} \hbar \omega_c)$ as a guide of the eyes.
    The black solid curve in (b) represents the insulating $\sigma_{xy}=0$ regime, which is the same as $\nu(1, \Gamma)$ at $T=0$.
    (c-d) are the corresponding results at finite temperature $T/\hbar \omega_c = 0.05$. The plateau width is defined as the distance between two edges of the peaks of $d\sigma_{xy}/d\nu$ [peak widths are the shaded area in (c)].}
    \label{fig:plateau_width}
\end{figure*}
For the case of $\Gamma > \hbar \omega_c$, LLs strongly overlap with each other. 
In this case, if the delocalized state still exists in the center of the LLs, then the DOS of other LLs overlaps with $E_N$ such that the DOS near $E_N$ increases by a factor of $\sim \Gamma/\hbar \omega_c$. 
As a result, electrons with Fermi level $\abs{E_F - E_N} \simeq \hbar \omega_c \Delta \nu_p \le T$ can be thermally activated, giving 
\begin{align}
    \Delta \nu_p = T/\hbar \omega_c, \qif T< \hbar \omega_c < \Gamma.
\end{align}
The corresponding plateau width $\Delta \nu \approx 1-\Delta \nu_p$ then remains a constant independent of disorder, since the relevant energy scale competing with temperature in this situation is the LL separation or the cyclotron energy..
The physical picture for this scenario is shown in Fig.~\ref{fig:LL_DOS}(b).
For temperatures higher than $\hbar \omega_c$, the thermal averaging should completely suppress the plateau.
The deviation of $\sigma_{xy}$ from the Drude result $\sigma_{xy}^{\mathrm{D}} = -e n_e c/B$ is exponentially small in this high-temperature ($>$ cyclotron energy) situation, with a factor of $\mathcal{F}(x=T/\hbar \omega_c) = x/\sinh{x}$~\cite{Polyakov:2003}.
No IQHE would manifest in this case with magnetotransport following the Drude theory with very small corrections.

For an even larger disorder $\Gamma > N\hbar \omega_c$, with $N$ distinct LLs overlapping strongly because of the highly broadened DOS, localized states in the $N$-th LL start accumulating below the lower LL that contribute to a long DOS tail shown in Fig.~\ref{fig:LL_DOS}(c).
Such a long localized DOS tail effectively increases the filling factor related to the center of LLs.
In other words, the $N$-th LL floats up to a filling factor greater than $N$ if the disorder is sufficiently large $\Gamma > N\hbar \omega_c$.
Since there is no LL below the lowest LL, the overlapped DOS near $E_N$ is smaller than the case shown in Fig.~\ref{fig:LL_DOS}(b), and $\Delta \nu_p$ decreases again as $\Gamma$ increases.
As a result, the finite-temperature plateau width for $\Gamma > N\hbar \omega_c$ should be roughly the same as the zero-temperature plateau width $\Delta \nu(T=0)$ where floating physics dominates.
Thus, at any finite $T$, sufficiently large disorder would suppress the IQHE and shrink the plateau because eventually the floating physics would dominate.  Since for $T=0$, any finite disorder is by definition very strong disorder, the plateaus always must shrink with increasing disorder at $T=0$ (as we already found in Sec.~\ref{sec:tight-binding}).

There is another logical possibility that, for sufficiently large disorder and low temperature, the delocalized states at the center of LLs are pushed to arbitrarily high 
densities 
by disorder,
so that the $\sigma_{xy}$ plateau disappears (as IQHE itself is destroyed by disorder)
and the whole system becomes an insulator (with all states localized) where both $\sigma_{xx}$ and $\sigma_{xy}$ become equal to zero.
This metal-insulator transition should occur for $\delta U > E_F$, where the electrons break into puddles separated from each other by large disorder potential barriers.
Using the expression of $\delta U$ Eq.~\eqref{eq:remote_potential_square}, we obtain the critical 2DEG density of this MIT as $n_c \sim \sqrt{n_i}/d$, which agrees with the results in Refs.~\cite{Efros:19881019,Pikus:1989,Tracy:2009,huang2023understanding}.
An equivalent way of stating this is that, if the $B=0$ system is already very strongly localized into well-separated puddles due to the very large disorder, then the application of a magnetic field should not lead to the delocalization at the LL centers necessary for the manifestation of IQHE. 
One relevant point to note here is that such a puddle-induced percolation localization transition necessarily involves macroscopic inhomogeneities because the system breaks into mutually insulating regimes, and it is possible that IQHE in a macroscopically inhomogeneous situation is not meaningful.

\begin{figure*}
    \centering
    \includegraphics[width=\linewidth]{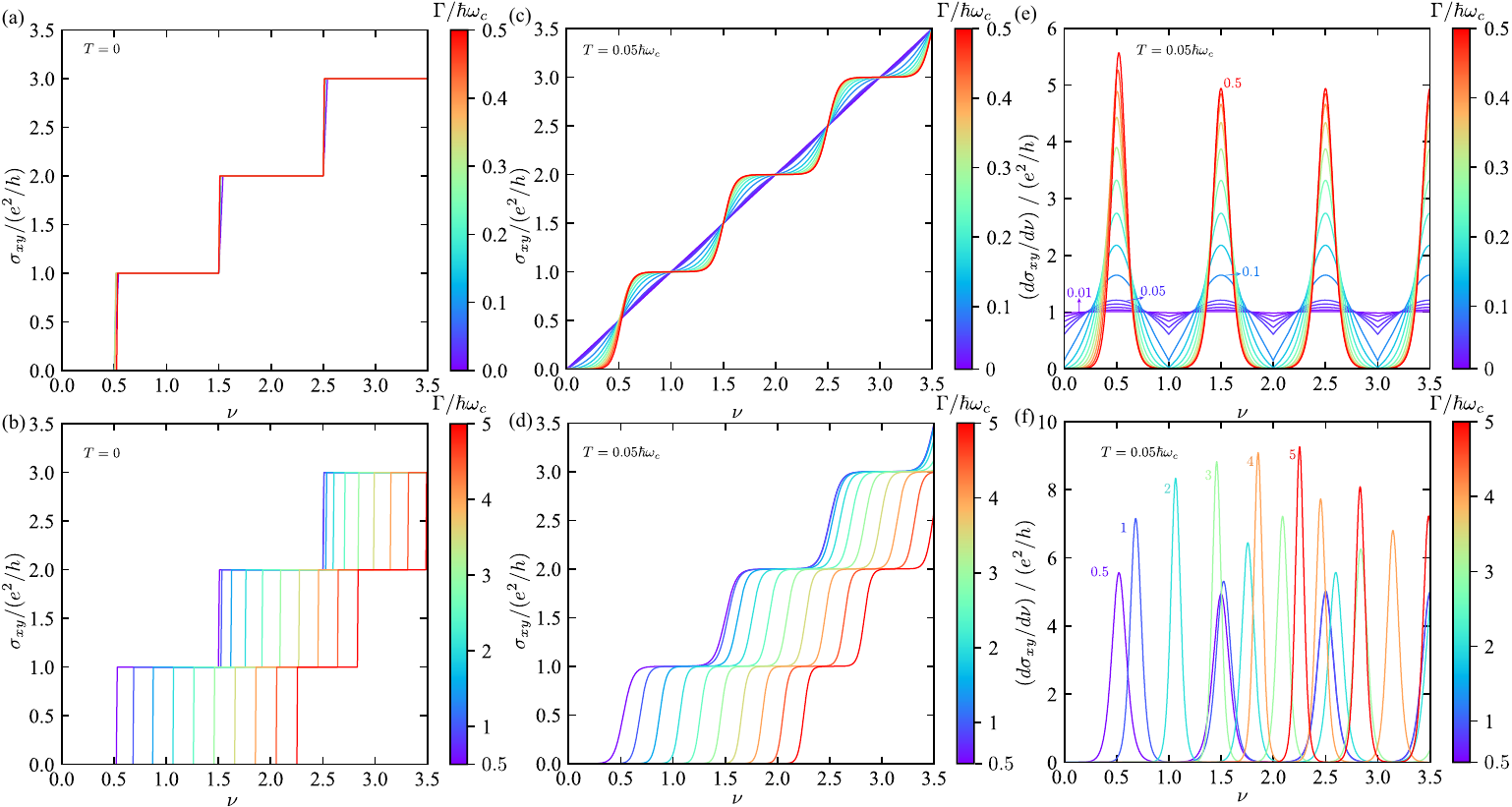}
    \caption{$T=0$ result of $\sigma_{xy}(\nu)$ for (a) weak disorder $\Gamma/\hbar\omega_c \in [0,0.5]$, and (b) strong disorder $\Gamma/\hbar\omega_c \in [0.5,5]$. 
    (c-d) are the corresponding results of $\sigma_{xy}(\nu)$ at a finite temperature $T/\hbar \omega_c=0.05$. (e-f) are the corresponding results of $d\sigma_{xy}/d\nu$ at a finite temperature $T/\hbar \omega_c=0.05$. The colored numbers label the value of $\Gamma/\hbar \omega_c$ used in the calculation.
    }
    \label{fig:sigmaxy_nu}
\end{figure*}

Given a disorder broadened DOS (i.e.\ Eq.~\ref{eq:LL_DOS_gaussian}), the corresponding filling factor at a chemical potential $\mu$ is given by
\begin{align}
    \nu(\mu,\Gamma) &= \sum_N \int_{-\infty}^{+\infty} d\varepsilon \Theta(\mu - \varepsilon) \frac{1}{\sqrt{2\pi \Gamma^2}} e^{-\frac{(\varepsilon-E_N)^2}{2\Gamma^2}} \\
    &= \frac{1}{2}\sum_N \mathrm{erfc}\qty(\frac{E_N - \mu}{\sqrt{2} \Gamma}). \label{eq:nu_N_T0}
\end{align}
The asymptotic behavior of $\nu(E_N,\Gamma)$ in the limit of small/large disorder is given by [cf.\ Fig.~\ref{fig:plateau_width} (a)]
\begin{align}
    \nu(E_N,\Gamma) = 
    \begin{cases}
         N-1/2, &\qif \Gamma \ll E_N, \\
         \Gamma/(\sqrt{2\pi} \hbar \omega_c), &\qif \Gamma \gg E_N.
    \end{cases}
\end{align}
The Hall conductivity at a chemical potential $\mu$ is given by
\begin{align}
    \sigma_{xy}(\mu) = \frac{e^2}{h} \sum_N \Theta(\mu - E_N),
\end{align}
with steps of $e^2/h$ at $\mu = E_N$. 
The corresponding plateau width that corresponds to the QH plateau $\sigma_{xy} = N e^2/h$ is given by
\begin{align}\label{eq:plateau_width_T0}
    &\Delta \nu (N,\Gamma,T=0) = \nu(E_{N+1},\Gamma) - \nu(E_{N},\Gamma), \\
    &= \frac{1}{2} \mathrm{erfc}\qty(-\frac{E_N}{\sqrt{2}\Gamma}), \\
    &\approx 
    \begin{cases}
        1 - \frac{\Gamma}{\sqrt{2\pi} E_N} e^{-E_N^2/2\Gamma^2}, &\qif \Gamma \ll E_N, \\
        \frac{1}{2} + \frac{E_N}{\sqrt{2\pi} \Gamma}, &\qif \Gamma \gg E_N.
    \end{cases}
\end{align}
We can parametrically plot $\sigma_{xy}(\mu)$ at $T=0$ as a function of $\nu(\mu,\Gamma)$ for different $\Gamma/\hbar \omega_c \in [0,5]$ as shown in Figs.~\ref{fig:sigmaxy_nu} (a-b). (The results for a larger range of disorder $\Gamma/\hbar \omega_c \in [0,10]$ and filling factors are presented in Fig.~\ref{fig:sigmaxy_nu_7LL} of Appendix~\ref{appendix:sigma_xy}).
In Figs.~\ref{fig:plateau_width} (a-b), the position of the $T=0$ plateau transition, $\nu(E_N,\Gamma)$, corresponding to the jump of $\sigma_{xy}/(e^2/h)$ from $N-1$ to $N$, is illustrated as a function of $\Gamma$.
Note that this result, which is qualitatively consistent with the exact finite-size numerical results for short-range disorder presented in Fig.~\ref{fig:peak-positions}, is quite distinct from the floating of Landau levels envisioned by Khmelnitzkii~\cite{khmelnitskii1984} and discussed by Laughlin~\cite{laughlin1984}. 
The latter results apply to the limit $E_N\gg \Gamma$ where they conjecture a renormalization of $\Delta\nu\propto [1+(\Gamma/\omega_c)^{2}]$, which is greater than the value of unity found here in this limit. 
One possible origin of this discrepancy is the use of the percolation model or classical limit in deriving our conclusions in this section. 
In this limit, the magnetic field can efficiently lead to the localized phases shown in between the lines (i.e., extended percolating states at the LL centers) in  Fig.~\ref{fig:plateau_width}(a) when the cyclotron radius $R_c$ is smaller than the disorder correlation length $d$ i.e. for fillings $\nu< \nu_c= m  \omega_c d^2/2\hbar$. Beyond this range of filling in Fig.~\ref{fig:plateau_width}(a) and at large disorder broadening $\Gamma$, the semiclassical 2DEG system is expected to be a diffusive metal so that the Landau levels shown in the figure with the corresponding quantum Hall plateaus are eliminated.
The black dashed line in Fig.~\ref{fig:plateau_width}(a) that delineates the trivial localized states for $\nu>\nu_c$ becomes the classical percolation transition line ($E_F \sim \Gamma$). 
Thus, within the semiclassical approximation, there are no LLs or quantum Hall plateaus for disorder $\Gamma>\Gamma_c=(2\pi)^{1/2}\hbar\omega_c\nu_c$.  However, it is believed that coherent back scattering that is ignored in the semiclassical limit leads to a unitary symmetry class Anderson localized insulator~\cite{hikami1981anderson} for even $\nu>\nu_c$. 
In this case, the topological transitions~\cite{pruisken1984} between the different localized regions in Fig.~\ref{fig:plateau_width}(a) would need to continue beyond $\nu>\nu_c$ along trajectories that are not described by the percolation model. Whether these LLs are destroyed above some critical value of disorder such as $\Gamma_c$ depends on the trajectory of the LLs at large $\Gamma$ and is therefore unfortunately beyond the scope of this work. 
In particular, the percolation model becomes increasingly inaccurate for high LLs.  There is also the possibility that the speculative floating  conjecture in Refs.~\cite{khmelnitskii1984,laughlin1984}, which was made with the explicit aim of reconciling the necessary existence of finite field extended states for IQHE with the established Anderson localization of zero-field 2D orthogonal class, being completely heuristic, is not quite accurate in terms of its details.  
We clearly see floating of the high field extended states with increasing $\Gamma/\hbar\omega_c$ in both our exact numerical results for short-range disorder in Sec.~\ref{sec:tight-binding} and for our percolation model for long range slowly varying disorder in Sec. ~\ref{sec:percolation_model}.

\begin{figure}
    \centering
    \includegraphics[width=0.9\linewidth]{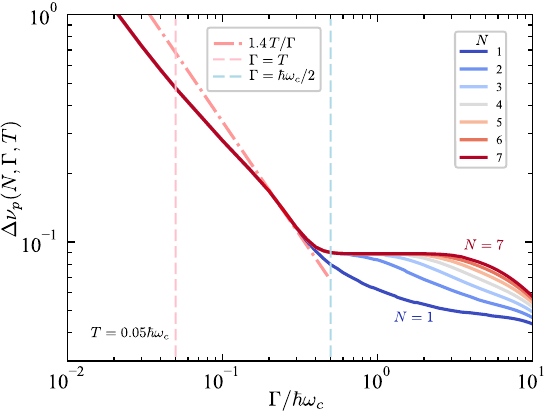}
    \caption{$T/\hbar \omega_c=0.05$ result of the half width $\Delta \nu_p$ of the peaks in $d\sigma_{xy}/d\nu$ as a function of disorder $\Gamma/\hbar\omega_c \in (0.01,10)$. For small value of $\Gamma/\hbar\omega_c < 0.15$ where LLs are overlapping due to strong thermal broadening, quadratic extrapolation is used to obtain the half width. The red dot-dashed line represents $1.4 T/\Gamma$ as a guide of the eyes.}
    \label{fig:half_width}
\end{figure}

Next, we generalize the above results to finite temperatures.
The filling factor at a finite $T$ is given by
\begin{align}
    \nu(\mu,\Gamma,T) = \sum_n \int_{-\infty}^{+\infty} d\varepsilon f\qty(\frac{\varepsilon - \mu}{T}) \frac{1}{\sqrt{2\pi \Gamma^2}} e^{-\frac{(\varepsilon-E_N)^2}{2\Gamma^2}},
\end{align}
where $f(x) = (e^x + 1)^{-1}$ is the Fermi-Dirac distribution function.
Because of the particle hole symmetry at $\mu=E_N$, the finite temperature filling factor at $\mu=E_N$ remains the same as the $T=0$ result
\begin{align}
    \nu(E_N,\Gamma,T) = \nu(E_N,\Gamma,T=0).
\end{align}
The Hall conductivity at finite $T$ is given by
\begin{align}
    \sigma_{xy}(\mu,T) = \frac{e^2}{h} \sum_N \qty[1- f\qty(\frac{\mu - E_N}{T})].
\end{align}
Apparently, at high temperatures $T>\hbar\omega_c$, the thermal broadening of the plateau transition completely destroys the plateau and $\sigma_{xy}$ becomes a straight line in $\nu$.
Indeed, no IQHE is ever experimentally reported for $T \sim \hbar\omega_c$.
Therefore, we focus on the low temperature regime where $T<\hbar \omega_c$.
The results of $\sigma_{xy}(\mu,T)$ as a function of $\nu(\mu,\Gamma,T)$ at $T/\hbar \omega_c = 0.05$ is shown in Figs.~\ref{fig:sigmaxy_nu}(c-d).
Experimentally, the width of this delocalized regime $\Delta \nu_p$ that corresponds to the peak of $\sigma_{xx}$ may be characterized by the slope of $\rho_{xy}$ as a function of the filling factor $\nu$, or $\max(d \rho_{xy}/d\nu)$, and $\Delta \nu_p \approx (h/e^2)\max(d \rho_{xy}/d\nu)^{-1}$~\cite{Tsui:1988,Tsui:1992,Klitzing:1991,Klitzing:1992,Dolgopolov:1991}.
We adopt this method to numerically compute $\Delta \nu_p$ and the corresponding plateau width $\Delta \nu$.
For example, to characterize the plateau quality at $T/\hbar \omega_c = 0.05$, the corresponding derivative $d\sigma_{xy}/d\nu$ is plotted in Figs.~\ref{fig:sigmaxy_nu} (e-f).
In Fig.~\ref{fig:half_width}, the peak of the derivative $d\sigma_{xy}/d\nu$ occurs at $\nu(N,\Gamma) = \nu(\mu=E_N,\Gamma)$.
The peak width, denoted by $\Delta\nu_p(n,\Gamma,T) = (\nu_{+}-\nu_{-})/2$, is characterized by the points $\nu_{\pm}$ where the derivative $d\sigma_{xy}(\nu_{\pm})/d\nu$ equals half of the peak's magnitude for $\sigma_{xy}$.
In situations where $d\sigma_{xy}/d\nu$ is sufficiently flat so that its minimum exceeds half the peak value, we perform a quadratic extrapolation in the vicinity of the peak. 
We then determine the positions of $\nu_{\pm}$ where the derivative obtained from this extrapolation equals half the peak value.
We find that $\Delta \nu_p \sim T/\Gamma$ is inversely proportional to $\Gamma$ for small disorder $\Gamma < \hbar \omega_c/2$, and approaches a constant $\Delta \nu_p \sim 2T/\hbar \omega_c$ for intermediate disorder $(\hbar \omega_c/2)<\Gamma < E_N/\sqrt{2\pi}$. 
The higher the LL index $N$, the longer is this flat regime of $\Delta \nu$.
For strong disorder $\Gamma > E_N/\sqrt{2\pi}$, $\Delta \nu_p$ decreases again when the floating of the peak position $\nu(N,\Gamma,T)$ commences.
Figure~\ref{fig:plateau_width}(c-d) illustrates the plateau transition filling factor, defined as the peak position $\nu(N,\Gamma,T)$ of $d\sigma_{xy}/d\nu$, with the transition broadening $\Delta\nu_p$ depicted as a shaded area.

Figure~\ref{fig:sigmaxy_nu}(c) shows that for low disorder such that $\Gamma < T$, the thermal activation completely smears out the plateau so that $\sigma_{xy}$ becomes a straight line linear in $\nu$, and the plateau width $\Delta \nu$ vanishes.
This appears to be an intriguing result implying that it is much more difficult to see IQHE quantization in high quality (i.e. small $\Gamma$) samples as one must go to much lower temperatures ($<\Gamma$) for the plateau to manifest.  
There is strong experimental support for this finding of ours in ultra high mobility 2D GaAs samples where extreme low temperatures ($< 10$ mK) are necessary to obtain IQHE quantization.~\cite{Myers:2021,Csathy_ultrahigh:2024,Myers:2024} 
The fact that increasing disorder may stabilize IQHE plateus were pointed out early in the experimental development of the subject.~\cite{Stormer:198232} 
We emphasize again that at $T=0$, however, the plateau width is maximal (i.e. unity, going entirely from one LL peak to the next at LL centers) for infinitesimal disorder, and increasing disorder can only suppress the plateau width, but experiments do not live at $T=0$.
For intermediate disorder $T<\Gamma<\hbar\omega_c/2$, the IQHE plateau starts to develop, with the width of the plateau transition $\Delta \nu_p \approx T/\Gamma$, so the transition is shaper as $\Gamma$ increases and we obtain a better plateau.
For higher disorder $(\hbar\omega_c/2)<\Gamma<E_N/\sqrt{2\pi}$, the plateau width $\Delta \nu \approx 1-\Delta \nu_p$ saturates with a fixed broadening for the plateau transition $\Delta \nu_p \approx 2T/\hbar \omega_c$.
For an even higher disorder $\Gamma>E_N/\sqrt{2\pi}$, the plateau width $\Delta \nu \approx 1/2 + N\hbar\omega_c/(\sqrt{2\pi} \Gamma)$ decreases as $\Gamma$ increases due to floating.
The insulating regime, characterized by both $\sigma_{xx} = 0$ and $\sigma_{xy} = 0$, extends over low filling factors, ranging from $\nu = 0$ to $\nu = \Gamma/(\sqrt{2\pi} \hbar \omega_c)$.
The percolation model predicts that the plateau width eventually saturates at $\Delta \nu = 1/2$ for $\Gamma \gg E_N$, because there are no LLs for negative energies. 
(If there were LLs at all integer enerigies $E_N/\hbar \omega_c=0, \pm1, \pm2, \dots$, then there would be no floating and $\Delta \nu(T=0)$ is always 1. In reality, all negative LLs, i.e., half of integer LLs are missing so $\Delta \nu$ saturates to $1/2$.)
However, since the lowest LL floats up to a filling factor of approximately $\nu \approx \Gamma/(\sqrt{2\pi} \hbar \omega_c)$, observing a quantized plateau in such strong disorder becomes challenging. 
This is because the localization condition in the percolation model requires $\nu < E_{Fc}/\hbar \omega_c$ [cf.\ Eq.~\eqref{eq:E_Fc}]. For $\nu > E_{Fc}/\hbar \omega_c$, the diffusion of the electron orbit guiding center becomes strong enough for electrons to traverse the entire disorder landscape, and the system transitions into a diffusive metal where both $\sigma_{xx}$ and $\sigma_{xy}$ follow the classical Drude formula-- this is the semiclassical magnetotransport physics studied extensively experimentally in weak magnetic fields where Landau quantization is unimportant and IQHE does not manifest.

\begin{figure}
    \centering
    \includegraphics[width=\linewidth]{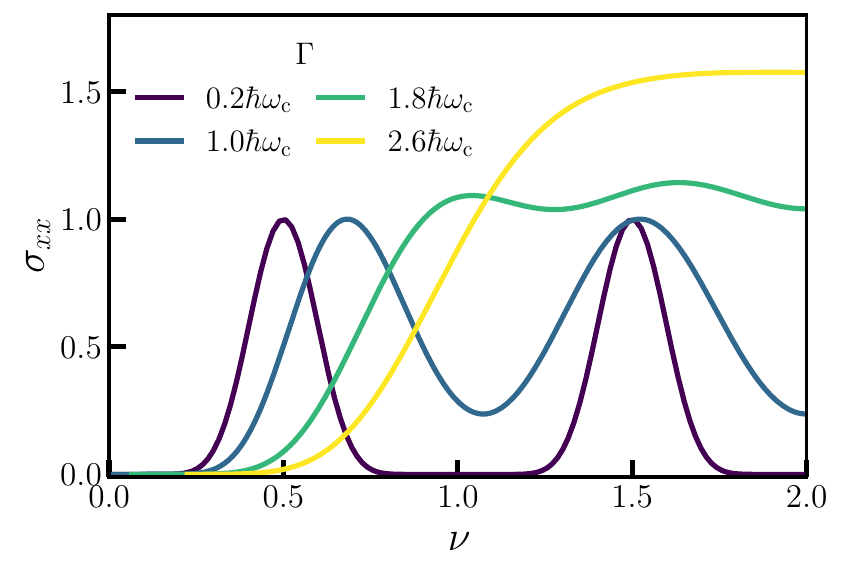}
    \caption{Conductance $\sigma_{xx}$ as a function of filling $\nu$ (compared to Fig.~\ref{fig:floating}) in the percolation model (i.e. Eq.~\ref{eq:sigmaxxperc})
for a finite system $L=40 l_B$ with increasing disorder $\Gamma$. The conductance shows well-defined peaks and plateaus at low disorder, which shift up (i.e. floating) as disorder $\Gamma$ increases. Concurrently the plateaus shrink and eventually at large $\Gamma$ the pleateau structure disappears. However, as seen from Eq.~\ref{eq:sigmaxxperc} increasing the system-size increases the scale of $\Gamma$, so that the plateaus are always sharp in the thermodynamic limit.}    \label{fig:percolationsigma}
\end{figure}

To understand the transition of the $T=0$ longitudinal conductance $\sigma_{xx}$ from peak-like to step-like behavior  as a function of disorder 
as seen in Fig.~\ref{fig:floating} from the percolation model, we need to consider a finite system size $L$. In the long-range 
disorder potential case i.e.\ $d\gg R_c$, the different Landau levels remain approximately decoupled because of the separation of length 
scales and energies of semi-classical orbits. As a result, the network model~\cite{huckestein1995,evers2008,chalker1988}, which is a generalized quantum lattice version of  the simpler percolation model, is a reasonable description of the transport properties even in the quantum regime. 
(We refer to the reader to the literature~\cite{KRAMER2005211,huckestein1995,evers2008,chalker1988} for the details on the network model.)
For the purpose of applying the results from the network model, we will need to connect the energy $\varepsilon$ to the parameter $x$ used to describe distance to the quantum critical point~\cite{dresselhaus2022} through the relation $x\sim \varepsilon/\Gamma$. Similarly the dimensionless length  of the network model is set by $L/l_B$. Therefore, combining the conductance from each of the LLs, the longitudinal conductance as a function of $L$ is given by 
\begin{align}
    \sigma_{xx}(\varepsilon=\mu)=\sum_{N=1}^\infty F(\Gamma^{-1}(\varepsilon-E_N)(L/l_B)^{1/\alpha^*}),\label{eq:sigmaxxperc}
\end{align} where $\alpha^*=2.609$~\cite{dresselhaus2022} 
, and we will assume for definiteness that the universal scaling function for the conductance~\cite{dresselhaus2022}
can be approximated by $F(x)=e^{-x^2/2}$. 
(This assumption does not affect our conclusion, and changing the function to some other reasonable form produces similar results.)
Combining this equation with Eq.~\ref{eq:nu_N_T0} leads to an equation for $\sigma_{xx}(\nu,\Gamma)$ shown in Fig.~\ref{fig:percolationsigma} that can be compared to the results in Fig.~\ref{fig:floating}.
In Fig.~\ref{fig:floating}, we take the limit $R_c/d\to0$, where diffusion vanishes and transport near the plateau transition is dominated by drift along the classical percolation path, yielding a conductance quantum independent of the LL index. (On the other hand, diffusion contributes to the numerical results shown in Figs.~\ref{fig:floating}, \ref{fig:finiteT}. This leads to an increase of $\sigma_{xx}$ with $N$, since the diffusion length scales with $R_c$, resulting in a diffusion constant $D \propto R_c^2/\tau \propto N$~\cite{Ando:1974,CZS:1994,Raikh:1993}.)
Specifically, the calculated conductance profile for the network model in Fig.~\ref{fig:percolationsigma} depends on the disorder broadening $\Gamma$ 
at $T=0$ in a way that is qualitatively similar to Fig.~\ref{fig:floating}. In fact, similar to Fig.~\ref{fig:floating}, the low conductance dips that are associated with quantized Hall plateaus apparently disappear also in Fig.~\ref{fig:percolationsigma} above a critical strength of disorder. 
This feature in Fig.~\ref{fig:percolationsigma} is a result of the strong overlap of the peak widths from the sum in Eq.~\ref{eq:sigmaxxperc}.  Furthermore, Eq.~\ref{eq:sigmaxxperc} for $\sigma_{xx}(\varepsilon)$ suggests that the peak widths in Fig.~\ref{fig:floating} are proportional to $\Gamma \omega_c^{-1} (L/l_B)^{-1/\alpha^*}$ and go to zero as the system size $L$ approaches the thermodynamic limit. 
In other words, the finite system size $L$ represents an effective temperature that broadens the conductivity peak and smears the plateau transition.
This would suggest that in the thermodynamic limit the quantum Hall transitions represented by the peaks in Fig.~\ref{fig:floating}, including the strong disorder regime where the quantized Hall plateaus have apparently disappeared, become sharp transitions with quantized plateaus and follow the floating phase boundaries shown in Fig.~\ref{fig:plateau_width}(a).
Thus, within this picture, there is no direct $T=0$ transition from the quantum Hall phase to a completely localized system (with no quantum Hall plateaus) except in the limit of $\omega_c/\Gamma$ approaching zero (which is equivalent to the zero field limit).
It  should be noted, though, that the delocalization at strong disorder seen in Fig.~\ref{fig:floating} might be a consequence of phase coherence 
beyond the semiclassical limit, which breaks down when the cyclotron radius exceeds the disorder correlation length.  
At any finite temperature, however, the experimental situation would appear to be a disorder-induced complete suppression of quantum Hall plateaus at `large enough' disorder since temperature imposes a lower cut off on the observable plateau width.  This is, however, only a crossover and not a quantum phase transition.
The extent to which these conclusions survives beyond the network model approximation is beyond the scope of this work.

\subsection{Variable range hopping}\label{subsec:VRH}

Here we discuss the variable range hopping (VRH) transport theory near the center of a LL in the strong magnetic field limit $\Gamma \ll \hbar \omega_c$~\cite{Polyakov:1993}.
When the temperature $T$ is sufficiently low so that both activation to the mobility edge and excitation over potential barriers to nearby impurity sites become improbable, conduction is more efficient through VRH. 
This involves electron tunneling among states within an energy range approximately equal to $T$ around $E_F$, which is the most efficient transport process in strongly localized systems at the lowest temperatures~\cite{shklovskii1984,shklovskii:2024}. 

The discussion in the previous section about the delocalization induced by thermal activation ignores the contribution from electron hopping, assuming that as long as $E_F \neq E_N$ the localization length is small and remains a fixed constant such that hopping is restricted to localized states between nearest neighbors, which does not have any temperature dependence.
In realistic experiments, the localization length gradually diverges when $E_F$ is moving closer to $E_N$, so that hopping becomes more efficient than thermal activation, and it is possible to reduce the thermal activation energy by hopping to a longer distance than between nearest neighbors. 
It has been theoretically proposed~\cite{Trugman:1983} and experimentally shown~\cite{Klitzing:1992} that the localization length diverges as $\xi = \xi_0 \Delta \nu_p^{-\gamma} \propto \abs{E_F - E_N}^{-\gamma} $ with a critical exponent $\gamma = 7/3\approx 2.3$ as the LL center $E_N$ is approached.
$\xi_0$ is roughly equal to the localization length of integer filling or $\Delta \nu_p \approx 1/2$ (i.e., $E_F$ is at the center of the $\sigma_{xy}$ plateau where $\sigma_{xx}$ reaches the minimum).
In principle, the prefactor $\xi_0$ should depend on the magnetic field $B$, the strength of the disorder $\delta U$, and the correlation length of the disorder potential $d$~\cite{Fogler:1998,Polyakov:2003,Shklovskii:1982}.
For example, for the white-noise short-range random disorder potential in the strong magnetic field limit $\hbar \omega_c \gg \Gamma$, $\xi_0 \simeq R_c$ coincides with the classical cyclotron radius, and for the lowest LL $\xi_0 \simeq l_B$~\cite{Fogler:1998,Polyakov:2003}.
For smooth long-range disorder see more discussion in Ref.~\cite{Fogler:1998}.

Phenomenologically, the deviation of $\sigma_{xy}$ from the integer quantized plateau value $N e^2/h$ is related to the longitudinal conductivity $\delta \sigma_{xy} \propto \sigma_{xx}$~\cite{Ando:1974b}, where $\sigma_{xx}$ taking into account both thermal activation and electron hopping is given by
\begin{align}
    \sigma_{xx} = \sigma_0 \exp(-\frac{\varepsilon_0}{T} - \frac{r}{\xi}),
\end{align}
where $\varepsilon_0$ is the typical activation energy near the Fermi level which represents the impurity band width for electrons hopping, and $r$ is the typical hopping distance between states inside the impurity band of width $\varepsilon_0$.
Both $\varepsilon_0$ and $r$ are functions of $E_F$ and $T$. 
The prefactor, which represents the maximum longitudinal conductivity of the $N$-th LL, can be estimated using SCBA as $\sigma_0 \sim N (e^2/h) (\tau_q/\tau)$~\cite{Ando:1974,CZS:1994,huang:2020,fu:2020}.
(In principle, the prefactor $\sigma_0$ is also a function of $T$, but the temperature dependence is a power-law type, so we ignore it and only consider the temperature dependence in the exponent).
To reduce the total energy of a transport system in the presence of an external current source, $\sigma_{xx}$ should be optimized so that only the channel with high conductivity is conducting and all other more resistive channels are short-circuited.
This optimization between thermal activation and hopping leads to VRH.

\begin{figure}
    \centering
    \includegraphics[width=0.9\linewidth]{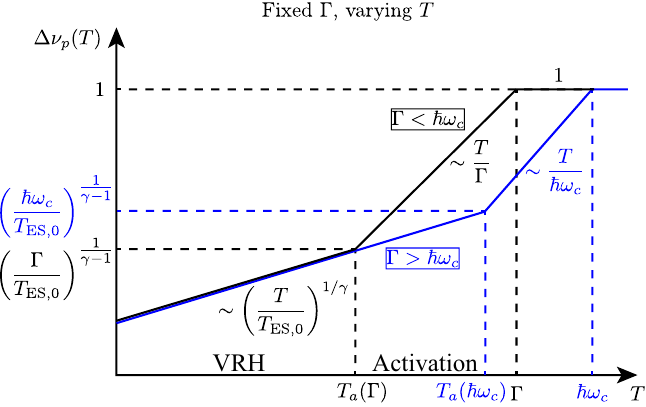}
    \caption{A schematic log-log plot of the range of filling factor that deviates from the quantized plateau $\Delta \nu(T)$ as a function of $T$. $\gamma = 7/3\approx 2.3$. At low temperatures $T<T_a$ [cf.\ Eq.~\eqref{eq:T_a}], delocalization is due to VRH; while at high temperatures $T>T_a$, delocalization is due to thermal activation. The plateau is completely destroyed at sufficiently large temperatures. Black curve represents the case for $\Gamma < \hbar \omega_c$, while the blue curve represents the case for $\Gamma > \hbar \omega_c$. By increasing $\Gamma$, the black curve continuously goes to the blue curve and saturates. In this plot, we assume $T_{\mathrm{ES},0}$ is the largest energy scale that does not depend on $\Gamma$ for simplicity.}
    \label{fig:delta_nu_T}
\end{figure}
Since the number of states per area inside $\varepsilon_0$ is given by
\begin{align}
    n(\varepsilon_0) = \int_{E_F-\varepsilon_0}^{E_F + \varepsilon_0} d\varepsilon g(\varepsilon),
\end{align}
the average hopping distance $r$ is related to $\varepsilon_0$ through $r = n(\varepsilon_0)^{-1/2}$.
For Mott VRH, the DOS near $E_F$ is assumed to be a constant~\cite{Mott:1969}; while for Efros-Shklovskii (ES) VRH, the DOS near $E_F$ has a Coulomb gap induced by the electron-electron interaction~\cite{shklovskii1984,shklovskii:2024}:
\begin{align}
    g(\varepsilon) = 
    \begin{cases}
        \kappa^2 \abs{\varepsilon-E_F}/e^4, &\text{ES VRH},\\
        g_0 \hbar \omega_c / 2\Gamma, & \text{Mott VRH}.
    \end{cases}
\end{align}
At $\abs{\varepsilon-E_F} = \varepsilon_g$, where
\begin{align}\label{eq:epsilon_g}
    \varepsilon_g = \qty(\frac{e^2}{\kappa l_B})^2\frac{1}{2\pi \Gamma},
\end{align}
the Coulomb gap DOS crosses over to the constant DOS.
Therefore, Mott DOS is valid if $\varepsilon_0 \gg \varepsilon_g$, and the Coulomb gap DOS is valid if $\varepsilon_0 < \varepsilon_g$.
For both cases, the number of states inside $\epsilon_0$ monotonically increases as $\epsilon_0$ increases. 
As a result, the typical hopping distance between impurities $r(\varepsilon_0)$ decreases as $\epsilon_0$ increases.
On the other hand, the thermal activation term $\varepsilon_0/T$ increases as $\epsilon_0$ increases, and the competition between electron hopping and thermal activation leads to the optimization of $\sigma_{xx}$ with respect to $\varepsilon_0$, such that 
\begin{align}
    \frac{\varepsilon_0}{T} = \frac{r(\varepsilon_0)}{\xi}.
\end{align}
The optimal energy band is given by
\begin{align}
   \varepsilon_0(T) = 
   \begin{cases}
        \sqrt{T e^2/\kappa \xi}, & \text{ES VRH}, \\
       \Gamma^{1/3} T^{2/3} (l_B/\xi)^{2/3}, & \text{Mott VRH}.
   \end{cases}
\end{align}
The optimal hopping distance is given by
\begin{align}
    r(T) = \xi 
    \begin{cases}
        \sqrt{e^2/\kappa \xi T}, & \text{ES VRH}, \\
       (\Gamma/T)^{1/3} (l_B/\xi)^{2/3}, & \text{Mott VRH}.
   \end{cases}
\end{align}
The corresponding conductivity reads
\begin{align}\label{eq:sigma_xx_VRH}
    \sigma_{xx}(T) = \sigma_0
    \begin{cases}
        \exp[-(T_{\mathrm{ES}}/T)^{1/2}], & \text{ES VRH}, \\
        \exp[-(T_{\mathrm{M}}/T)^{1/3}], & \text{Mott VRH},\\
    \end{cases}
\end{align}
where 
\begin{gather}
    T_{\mathrm{ES}} = \frac{e^2}{\kappa \xi} = \frac{e^2}{\kappa \xi_0} \Delta \nu_p^{\gamma}, \\
    T_{\mathrm{M}} = \Gamma (l_B/\xi)^2 = \Gamma (l_B/\xi_0)^2 \Delta \nu_p^{2\gamma}.
\end{gather}
Since $\xi = \xi_0 \Delta \nu_p^{-\gamma}$, the width of the conductivity peak is determined by equating $T$ with $T_{\mathrm{ES}}(\Delta \nu)$ or $T_{\mathrm{M}}(\Delta \nu)$, and we obtain
\begin{align}\label{eq:delta_nu_VRH}
    \Delta \nu_p = 
    \begin{cases}
        (T/T_{\mathrm{ES},0})^{1/\gamma}, & \text{ES VRH}, \\
        (T/T_{\mathrm{M},0})^{1/2\gamma}, & \text{Mott VRH},\\
    \end{cases}
\end{align}
where 
\begin{gather}
    T_{\mathrm{ES},0} = e^2/\kappa \xi_0, \\
    T_{\mathrm{M},0} = \Gamma (l_B/\xi_0)^2.
\end{gather}
The exponent predicted by ES VRH is $1/\gamma \approx 0.4$ agrees well with experiments for spin-split LLs~\cite{Tsui:1988,Tsui:1992,Klitzing:1991,Klitzing:1992,Dolgopolov:1991}.

Equating the VRH conductivity Eq.~\eqref{eq:sigma_xx_VRH} with the thermal activated conductivity $\sigma_{xx}=\sigma_0 \exp(-\abs{E_F - E_N}/T)$, we find the thermal activation regime corresponds to high temperatures 
\begin{align}\label{eq:T_activation}
    T> T_a = 
    \begin{cases}
        \Gamma^2/T_{\mathrm{ES},0} \Delta \nu_p^{2-\gamma}, & \text{for ES VRH}, \\
        \Gamma (\xi_0/l_B) \Delta \nu_p^{3/2-\gamma}, & \text{for Mott VRH}.
    \end{cases}
\end{align}
For relatively small level broadening $\Gamma < T_{\mathrm{ES},0},T_{\mathrm{M},0}$, substituting Eq.~\eqref{eq:delta_nu_VRH} into Eq.~\eqref{eq:T_activation} (or solving $\Gamma \delta \nu < T$), we obtain the crossover temperature between VRH and activation conductivity explicitly:
\begin{align}\label{eq:T_a}
    T_a = 
    \begin{cases}
        \Gamma (\Gamma/T_{\mathrm{ES},0})^{\tfrac{1}{\gamma - 1}}, & \text{if $\Gamma < T_{\mathrm{ES},0}$ for ES VRH}, \\
        \Gamma (\Gamma/T_{\mathrm{M},0})^{\tfrac{1}{2\gamma - 1}}, & \text{if $\Gamma < T_{\mathrm{M},0}$ for Mott VRH}.
    \end{cases}
\end{align}
and VRH dominates the low-temperature regime $T<T_a$ with $\Delta \nu_p$ given by Eq.~\eqref{eq:delta_nu_VRH}, while thermal activation behavior dominates $T_a<T<\Gamma$ regime with $\Delta \nu_p$ given by Eq.~\eqref{eq:delta_nu_activation}. 
When the temperature exceeds $T>\Gamma$, all states are readily excited to the middle of the LL, causing the plateau to vanish entirely.
For relatively large level broadening $T_{\mathrm{ES},0},T_{\mathrm{M},0}< \Gamma < \hbar \omega_c$, activation across the center of LL is more difficult, and VRH through an impurity band near $E_F$ is more effective and should dominate the entire low-temperature range $T<\Gamma$.

Experimentally, ES VRH near half-integer fillings in the quantum Hall effect is observed at low temperatures 10 mK $<T< 1$ K in both InGaAs/InP~\cite{Briggs:1983} and GaAs/GaAlAs~\cite{Ebert:1983} heterostructures, and activated behavior is observed at higher temperature $T> 1$ K~\cite{Klitzing:1985} around $\nu=2$, which indicates $\Gamma < T_{\mathrm{ES},0}$ in those experiments. 
Using the experimental parameters for GaAs~\cite{Ebert:1983,Klitzing:1985}, $B\simeq 7$ T, $m=0.067 m_0$, $\kappa=13$, $\mu = 5\times 10^5$ cm$^2$/Vs, we find $\hbar \omega_c = 12.5$ meV, $\Gamma \approx \sqrt{\hbar^2 \omega_c/\tau} = 1$ meV, $T_{\mathrm{ES},0} \approx 10$ meV if we use $\xi_0\approx l_B$ for the lowest LL. 
Using Eq.~\eqref{eq:T_a}, we find $T_a \approx 1$ K in reasonable agreement with the crossover temperature in experiments~\cite{Polyakov:1993}.

Finally, we comment on Mott VRH, which is only seen experimentally in a very narrow range of temperatures 1 K $< T<$ 2 K~\cite{Tsui:1982,Stormer:198232}.
This narrow temperature range puts the experimental claim into doubt.
For relatively large level broadening such that 
\begin{align}
    \Gamma > \Gamma_g = \frac{e^2}{\kappa l_B} \frac{\xi}{l_B}, \label{eq:Gamma_g}
\end{align}
ES VRH crosses over to Mott VRH at a temperature determined by $\varepsilon_0(T)=\varepsilon_g$ or 
\begin{align}
    T=T_g = \frac{T_{\mathrm{ES},0}^3}{T_{\mathrm{M},0}^2} = \qty(\frac{e^2}{\kappa l_B})^3 \frac{1}{\Gamma^2} \frac{\xi}{l_B}.
\end{align}
However, using the experimental parameters, the predicted $T_g\gg \hbar \omega_c$ is so large that Eq.~\eqref{eq:Gamma_g} cannot be satisfied, and ES VRH should dominate over Mott VRH for the entire low temperature regime $T<T_a$.
For small level broadening $\Gamma < \Gamma_g$, the DOS has the Coulomb gap form for the whole relevant energy range near $E_F$, and ES VRH always dominates over Mott VRH.

The results of $\Delta \nu_p(T)$ that combine the delocalization mechanisms of ES VRH and thermal activation are summarized in Fig.~\ref{fig:delta_nu_T}, providing a more complete quantitative picture for the temperature dependence of IQHE.  
The qualitative results are, however, similar to the ones we obtained using only activated transport without any  VRH transport (if the localization length $\xi_0$ decreases as disorder $\Gamma$ increases).

\section{Conclusion}
\label{sec:conclusion}
In this paper we have theoretically studied the effects of disorder and temperature on IQHE using a variety of analytical and numerical techniques addressing a series of conceptually and experimentally relevant questions: Does the IQHE plateau grow or shrink when disorder is increased keeping all other parameters fixed?
How does the IQHE plateau depend on temperature?
Is there a phase transition of IQHE to localization (with no IQHE plateaus) induced by increasing disorder?  Can disorder destroy the IQHE completely? Do extended states at the middle of each Landau level float up in energy as disorder increases?  How do disorder and temperature affect different Landau levels? What are the competing roles of various energy scales (e.g.\ disorder, temperature, Landau level separation, chemical potential) in IQHE?
Although the IQHE is a single-particle noninteracting problem,  the physics  is nevertheless challenging because of the nontrivial presence of multiple energy scales in the problem, some of which (cyclotron energy, short-range and/or long-range disorder) appear in the Hamiltonian and others (temperature, Fermi energy, system size) do not.  
In addition, the system is topological, and has an underlying Chern number at $T=0$ (but not at finite temperature), which the theory must incorporate nonperturbatively.
We emphasize that finite temperature always suppresses IQHE, and increasing temperature in a sample leads to a continuous decrease in the plateau width with IQHE eventually becoming unobservable for temperatures far above the cyclotron energy (but this is not a phase transition of any kind; it is simply a consequence of thermal excitations overcoming the energy gap protecting IQHE).

We find the answers to these questions to be subtle and nuanced, perhaps explaining why these questions have rarely been addressed in a comprehensive manner in the theoretical literature, and why, to the extent they have been, the answers are often contradictory.  The IQHE phenomenology depends intricately on several energy and length scales, making decisive answers to the above questions complicated and difficult.  
In particular, temperature plays a key role in the physics which has not been discussed much in the literature except in the context of (unknown) phenomenological inelastic scattering length cut off and the dynamical exponent in the plateau to plateau transition~\cite{Tsui:1988,Tsui:1992,Klitzing:1991,Klitzing:1992,Dolgopolov:1991,Pudalov:1990}, which we are not studying in the current work. 
We find that disorder and temperature compete in controlling the plateau width, and while at $T=0$, disorder always shrinks the plateaus, starting from the lowest Landau level, finite temperature competes with this process, leading to increasing disorder expanding the plateau width in some parameter regimes.  
Eventually, however, for large enough disorder, the plateaus shrink again with increasing disorder even at finite temperature.
Thus, the dependence of the IQHE plateau width on disorder is nonmonotonic depending on temperature (as well as the cyclotron energy).  

Similarly, the  localization leading to the suppression of the IQHE at $T=0$ begins always in the lowest Landau level starting with vanishing Landau level filling (i.e.\ when the chemical potential is in the low energy tail of the lowest Landau level) moving upward with increasing disorder. 
This is the 'floating' scenario, which we verify explicitly.
The extended states at the center of Landau levels move up in energy as disorder increases, albeit very slowly, eventually producing, in the limit of zero field or infinite disorder, the expected 2D zero-field orthogonal class localization with no IQHE, but there is no field-induced phase transition as the full localization leading to the vanishing of all IQHE plateaus happens only in the limit of the dimensionless parameter $\omega_\mathrm{c} \tau$ ($\sim \hbar \omega_c/\Gamma$) vanishing.  
The apparent experimental observation of the vanishing of IQHE for high Landau levels with decreasing magnetic field and/or increasing chemical potential is simply a result of finite temperatures---the IQHE plateaus should always reappear as temperature is lowered in high Landau levels, but with deceasing plateau width as $\Gamma/\hbar \omega_c$ increases.  
(Any actual observation of such IQHE in high LLs or for large disorder requires going to arbitrarily low temperatures, and may be practically impossible.)
Our direct numerical observation of floating of the extended states does not in any way depend on having an underlying lattice, as is necessary for the Chern number calculations~\cite{sheng1997,sheng1998,yang1999}, where the lattice induced Chern numbers with opposite signs in the electron and hole sectors annihilate each other with increasing disorder (or decreasing magnetic field) leading to floating.

We use several complementary techniques for our comprehensive analysis since each technique has its own unique limitations.  Our most obvious technique is a direct approach of solving the disordered Hamiltonian directly numerically.  We calculate $\sigma_{xx}$ directly as a function of disorder, LL filling and system size.  
We obtain the corresponding finite-$T$ results by convolving the exact $T=0$ results with the Fermi distribution function.  
These results decisively establish both the plateau shrinkage and the associated floating of the conductance peaks (coinciding with extended states in each LL) to higher energy, but the results suffer from the finite size limitations, making firm quantitative conclusions about the thermodynamic limit difficult.  
We therefore also use the percolation model for approximate analytical calculations in the thermodynamic limit, obtaining results qualitatively identical to the ones from the direct numerical simulation.  
The finite temperature  plateau width is nonmonotonic as a function of increasing disorder with the width increasing with disorder at first, but eventually decreasing when the disorder is very large.  
At $T=0$, disorder only suppresses the plateau width with increasing disorder.

There is an additional rather `trivial' mechanism affecting the IQHE plateau width, particularly in cleaner 2D samples, which we have not discussed at all.  This arises from the invariable presence of the fractional quantum Hall effect (FQHE) in lower Landau levels (most particularly, the LLL) in clean high-mobility 2D samples, which suppresses the IQHE plateau formation at low temperatures.  Since increasing disorder always suppresses FQHE, this new physics of the competition between IQHE and FQHE causes a seeming stabilization and enhancement of IQHE plateaus with increasing disorder as the FQHE at fractional fillings are systematically destroyed by disorder.  We do not discuss this effect because it is rather obvious, and arises from the disorder-induced suppression of FQHE and is therefore not an intrinsic IQHE phenomenon.  Also, our theory explicitly neglects electron-electron interaction, and thus FQHE is beyond the scope of our work.  We note, however, that this competition between FQHE and IQHE further enhances the IQHE plateau width with increasing disorder, perhaps occurring even at $T=0$.  This FQHE-IQHE competition also leads to a possible stabilization and enhancement of the IQHE plateau width in the lower Landau levels with increasing temperature (even at a fixed disorder) since the FQHE energy gap is typically much smaller than the IQHE gap, therefore causing the FQHE being suppressed much faster with increasing temperature than IQHE, again producing a thermal expansion of IQHE into the fractional filling regions of FQHE initially with increasing temperature.  Eventually, of course, at sufficiently high temperatures, IQHE itself disappears due to thermal excitations as discussed in the current work.

We find some results which seem surprising at first.  
For example, we find that at finite temperatures the IQHE is fragile for high-quality samples with little disorder as one must go to very low temperatures to see well-formed plateaus.  
In fact, this has already been experimentally reported that ultra-high-mobility 2D samples require going to 5 mK for the manifestation of good IQHE plateaus~\cite{Myers:2021,Csathy_ultrahigh:2024,Myers:2024}.
The fact that IQHE plateaus expand with increasing disorder in large regimes of parameters has been known for a long time, and our work explains this physics~\cite{Stormer:198232}.
We emphasize that a system with no disorder cannot manifest any IQHE since there is no mobility gap in pure systems and the plateaus are all of measure zero.  
IQHE necessarily requires disorder in order to convert the spectral gap inherent in the Landau level energy levels to a mobility gap so that the quantized plateaus show up.  Another nonobvious result of ours is that IQHE, in principle, manifests for all finite disorder, no matter how large.  
The plateaus at high disorder are very small, and unobservable, but there is no disorder-induced quantum phase transition from IQHE to a localized system with no IQHE.  IQHE disappears asymptotically when the parameter $\omega_c/\Gamma$ vanishes as the 2D system becomes a localized insulator at zero field (or infinite disorder).  
Alternatively, at any fixed $\omega_c/\Gamma$ (however small), we find IQHE to survive over some range of filling $\nu$. There is, however, a $T=0$ localization transition from a $N=1$ IQHE phase to a trivially insulating phase as disorder $\Gamma$ is increased while at fixed filling $\nu$. 
We emphasize that there is no direct disorder-induced quantum phase transition suppressing all IQHE (i.e.\ all the extended states at the centers of all Landau levels) for any finite disorder, but only floating of extended states starting at the lowest Landau level moving upward, which may systematically suppress IQHE with increasing disorder at progressively higher Landau levels, again starting at the lowest Landau level.  The fact that IQHE is suppressed at the lowest filling with the system entering a highly resistive localized insulating state has been known for a long time---the original 1980 discovery of IQHE did not manifest any IQHE in the lowest two spin- and valley-split Landau levels, similar to what can be seen for higher disorder in our Fig.~\ref{fig:floating} (and schematically represented in our Fig.~\ref{fig:phase_diagram}) where increasing disorder induces localization at the lowest filling.
Curiously, IQHE vanishes also in zero disorder ($\Gamma=0$) limit since the existence of the plateau requires a mobility gap which happens only in the presence of  (perhaps infinitesimal) disorder.  
Thus, IQHE vanishes for both zero and infinite disorder, but not for any intermediate disorder.

This brings up the interesting question on what happens if a magnetic field is applied to a nominally strongly localized highly disordered 2D system. Our work is consistent with such a strongly localized $B=0$ 2D system manifesting IQHE provided the applied field is strong enough, so that $\Gamma/\omega_c$ is no longer large.  
Of course, for very large disorder, it may be impossible to violate the $\Gamma/\omega_c \gg1$ condition, so the actual experimental observation of a strong field IQHE in a weak-field strongly localized 2D system may be impractical because the necessary field to make $\Gamma/\omega_c$ not very large is unachievable.

Our work has obvious experimental implications, but the presence of FQHE in real systems complicates the picture as we have no FQHE in our model of noninteracting electrons.  (The absence of spin in our spinless electron model, however, does not cause any complications, with the only caveat being that spin splitting due to Zeeman effect adds another energy scale to the Landau level separation. The same is true for any possible valley degeneracy as applicable to 2D Si systems.)  In lower Landau levels, FQHE and IQHE compete, and there is the possibility of competition also with the Wigner crystal phase which may also arise from interaction.  
The existence of five possible competing phases (IQHE, FQHE, Wigner crystal, Fermi liquid, localized) is simply too complex for any theory to comment on, but in moderately disordered samples (e.g.\ Si MOSFETs), where no FQHE is observed, our results should apply directly.  
In general, disorder has a strong effect at low Landau level filling, and the possibility cannot be ruled out that the disappearance of both FQHE and IQHE  at very low filling, as observed experimentally, may very well arise from disorder-induced localization causing floating (at least in some situations).  
Our work should stimulate new experiments to probe the role of disorder and temperature in IQHE focusing on the lower few Landau levels using tunable disorder and low temperatures.
\begin{acknowledgments}
S.Y.T.\@ thanks the Joint Quantum Institute at the University of Maryland for support through a JQI fellowship.
This work is also supported by the Laboratory for Physical Sciences through its continuous support of the Condensed Matter Theory Center at the University of Maryland.
\end{acknowledgments}
\section*{Data availability}
The data from the tight-binding calculations are openly available~\cite{data}.

\bibliography{bib.bib}

\appendix 
\section{Additional tight-binding details}
\label{sec:kwant_app}
\begin{figure}[h]
    \vspace{1em}
    \centering
    \begin{overpic}[width=1\linewidth]{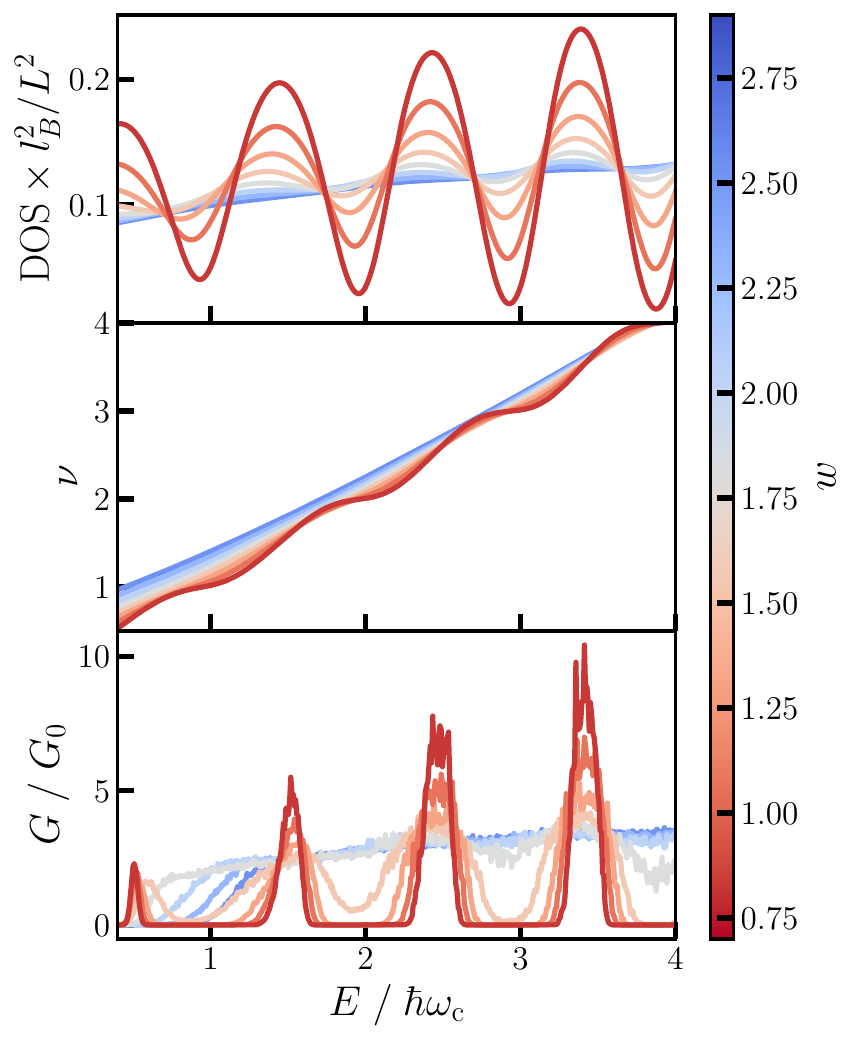}
    \put(15,94){(a)}
    \put(15,64.5){(b)}
    \put(15,35){(c)}
    \end{overpic}
    \caption{Data used to produce Fig.~\ref{fig:floating}. (a) Quantum oscillations in the density of states showing periodicity with the cyclotron frequency. (b) Filling $\nu$, calculated from the integrated density of states with Eq.~\ref{eq:filling}. (c) Transverse conductance $G$ as a function of $E$ for various disorder values. $L=800$.
    We use approximately $1200$ disorder realizations.
    }
    \label{fig:details}
\end{figure}

\begin{figure}[h]
    \vspace{1em}
    \centering
    \includegraphics[width=1\linewidth]{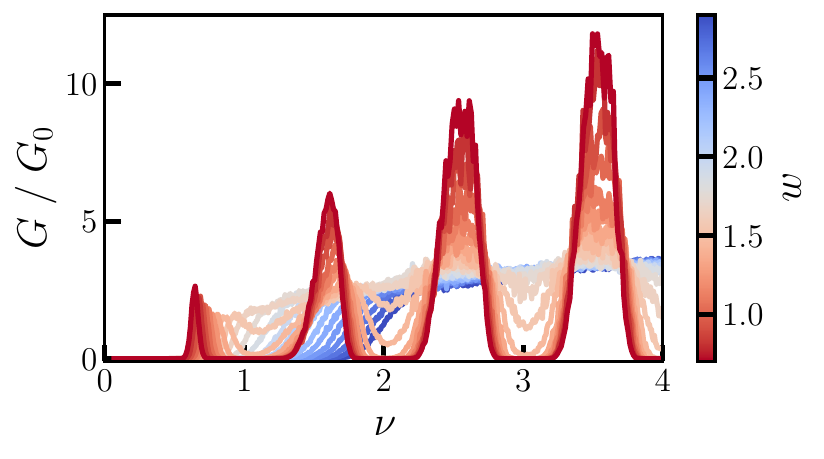}
    \caption{%
    Longitudinal conductance $G$ of the first two Landau levels as a function of filling $\nu$ with increasing disorder strength $w$, showing a wider range of fillings than Fig.~\ref{fig:floating}. %
    $L=800$, $l_B=4$. %
    We use approximately $1200$ disorder realizations for $\nu<2$ and $300$ for $\nu>2$. %
    }
    \label{fig:floating4}
\end{figure}

Calculating the filling $\nu$ at different energies amounts to integrating the global density of states. 
Since the kernel polynomial method finds the density of states in the form of a power series, finding the integral, and thus the cumulative density of states, is trivial. 
The density of states of the tight-binding model shows clear quantum oscillations, as shown in Fig.~\ref{fig:details}(a). 
This value can be integrated to get the filling
\begin{equation}
\label{eq:filling}
\nu(E) =  \frac{l_B^2}{2\pi L^2} \int_{-\infty}^E dE'\; \mathrm{DOS}(E').
\end{equation}
which is shown in Fig.~\ref{fig:details}(b).
We can compare this quantity with the conductance as a function of energy given in Fig.~\ref{fig:details}(c) to find the conductance as a function of filling $\nu$ in Fig.~\ref{fig:floating}. All four Landau levels are shown in Fig.~\ref{fig:floating4}. The upwards drift of the filling in Fig.~\ref{fig:details}(b) with increasing disorder results in the floating behavior discussed in Sec.~\ref{sec:floating}.

In calculating the density of states using the polynomial method, the maximal range of the Hamiltonian eigenvalues must be known exactly or polynomial fitting will suffer instabilities. Unlike Runge instabilities, the Lanczos method used here is numerically stable given that this range includes all of the eigenvalues. To ensure this is the case, the kernel polynomial method utilizes a tolerance parameter $\epsilon$ such that all eigenvalues are a distance $\Delta E / (2-\epsilon)$ from the average eigenvalue, where $\Delta E$ is the measured range (measured using the Lanczos algorithm). We use a value of $\epsilon=0.1$ which empirically resolves instabilities.
\begin{figure}[h]
    \centering
    \begin{overpic}[width=1\linewidth]{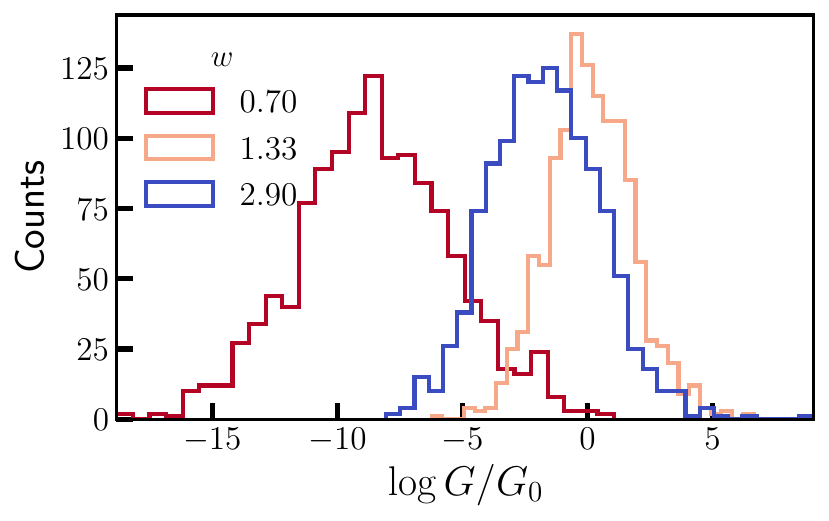}
    \end{overpic}
    \caption{Histogram of conductance measurements in $L=800$ system for different disorder strengths, demonstrating an approximate log-normal distribution. Measurements were taken at $E=1.28\omegac$.}
    \label{fig:histogram}
\end{figure}
As noted in the main text, the distribution of conductances is skewed, approximating a log-normal distribution~\cite{dresselhaus2022}. This distribution is demonstrated in Fig.~\ref{fig:histogram}.
\begin{figure}[h]
\vspace{4pt}
    \centering
    \includegraphics[width=1\linewidth]{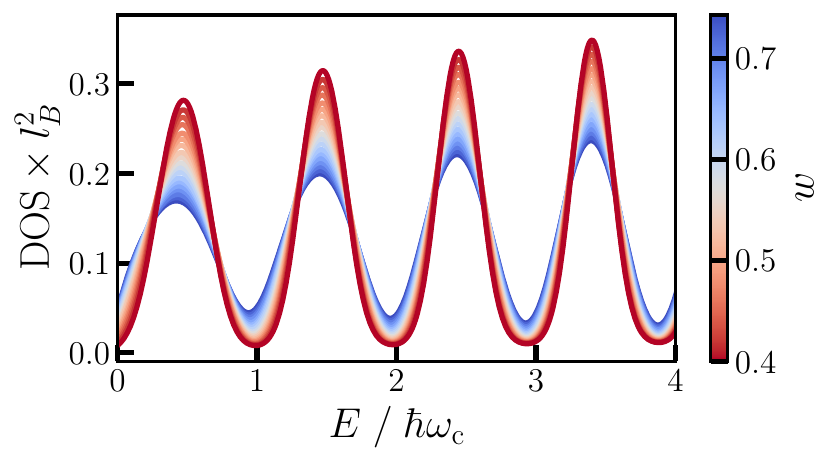}
    \caption{Density of states at $T=0.05\omega_c$, used to produce Fig.~\ref{fig:finiteT}. We use approximately $1200$ realizations}
    \label{fig:dos-finiteT}
\end{figure}

The increased temperature broadens the density of states, which results in the data shown in Fig.~\ref{fig:dos-finiteT}. The broadening $\Gamma$ remains smaller than the cyclotron frequency $\hbar \omega_\mathrm{c}$, indicating that disorder strengths $w$ between $0.4$ and $0.7$ lead to the thermally activated regime shown in Fig.~\ref{fig:plateau_width_schematic}.

\section{Anti-levitation}
\begin{figure}
    \centering
    \includegraphics[width=1\linewidth]{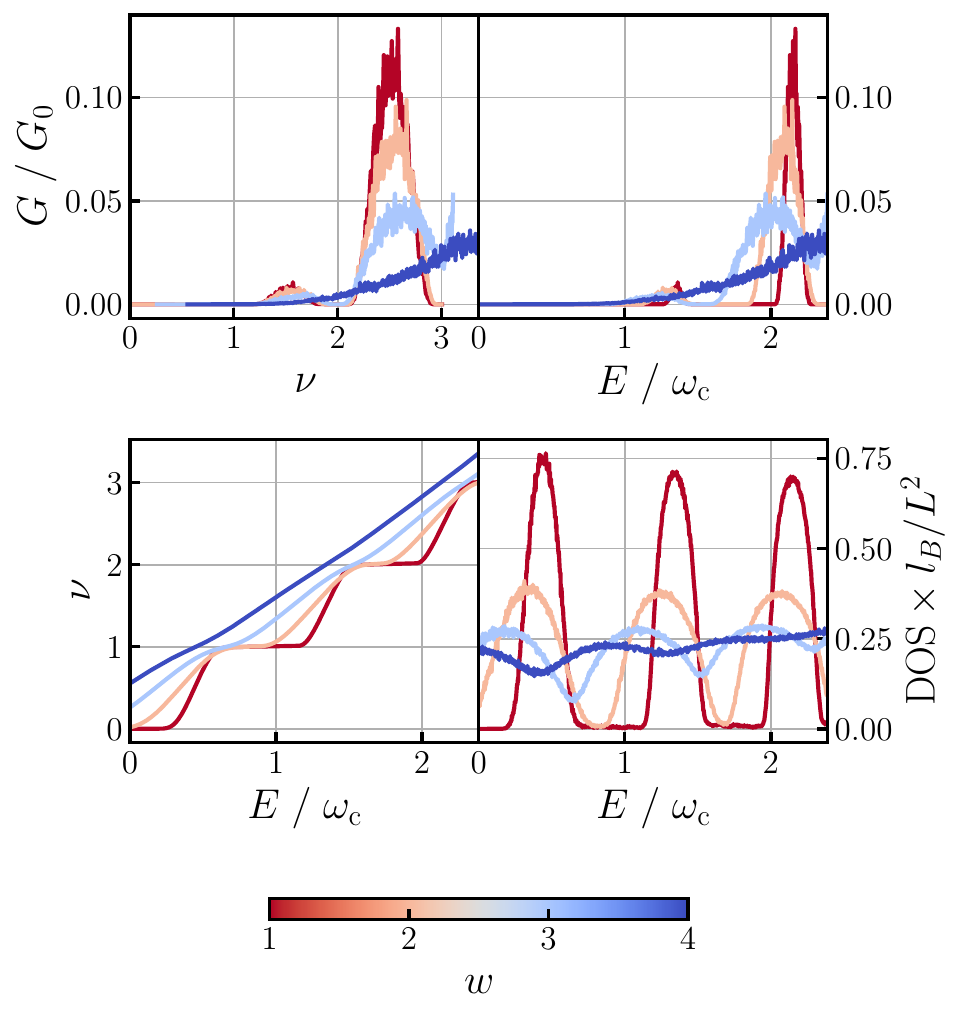}
    \caption{Anti-levitation of energy levels demonstrated in density of states and conductance. $l_\mathrm{B}=1.5a$. $L=400$. We use $500$ realizations.}
    \label{fig:antilevitation}
\end{figure}

In some existing studies~\cite{wang2014,pan2016}, the Landau levels seem to exhibit ``anti-levitation,'' where the extended states shift down in energy. This was seen experimentally in Ref.~\onlinecite{pan2016} and studied theoretically in Ref.~\onlinecite{wang2014}. We expect that this phenomenon is a lattice effect since in Ref.~\onlinecite{wang2014}, the magnetic length is on the order of the lattice constant. We verify this claim with our own calculation by increasing the magnetic field such that $l_\mathrm{B}=1.5a$. We show these results in Fig.~\ref{fig:antilevitation}. When $w\lesssim 3$ the energies of the DOS peaks decrease with increasing disorder.  Additionally, the conductance peaks seem to shift to lower energies before increasing at $w\approx 4$. However the floating of the cumulative density of states counteracts this effect and no anti-levitation is seen in the conductance as a function of filling. While this $l_B>a$ lattice effect explains the theoretical results in Ref.~\onlinecite{wang2014}, it cannot account for the experimental anti-levitation observed in Ref.~\onlinecite{pan2016}. The experiment is conducted at small magnetic field $B<0.3$ T where the magnetic length $l_B>48$ nm is much larger than the lattice constant of GaAs. This suggests that exchange-interaction-modified Zeeman splitting, which was not included in the numerical tight-binding calculation, may be important for explaining the observed anti-levitation behavior. A detailed exploration of this mechanism, however, is beyond the scope of the current paper.

\section{Hall conductivity in the percolation model for a larger range of disorder and filling factors}
\label{appendix:sigma_xy}
\begin{figure*}[t]
    \centering
    \includegraphics[width=\linewidth]{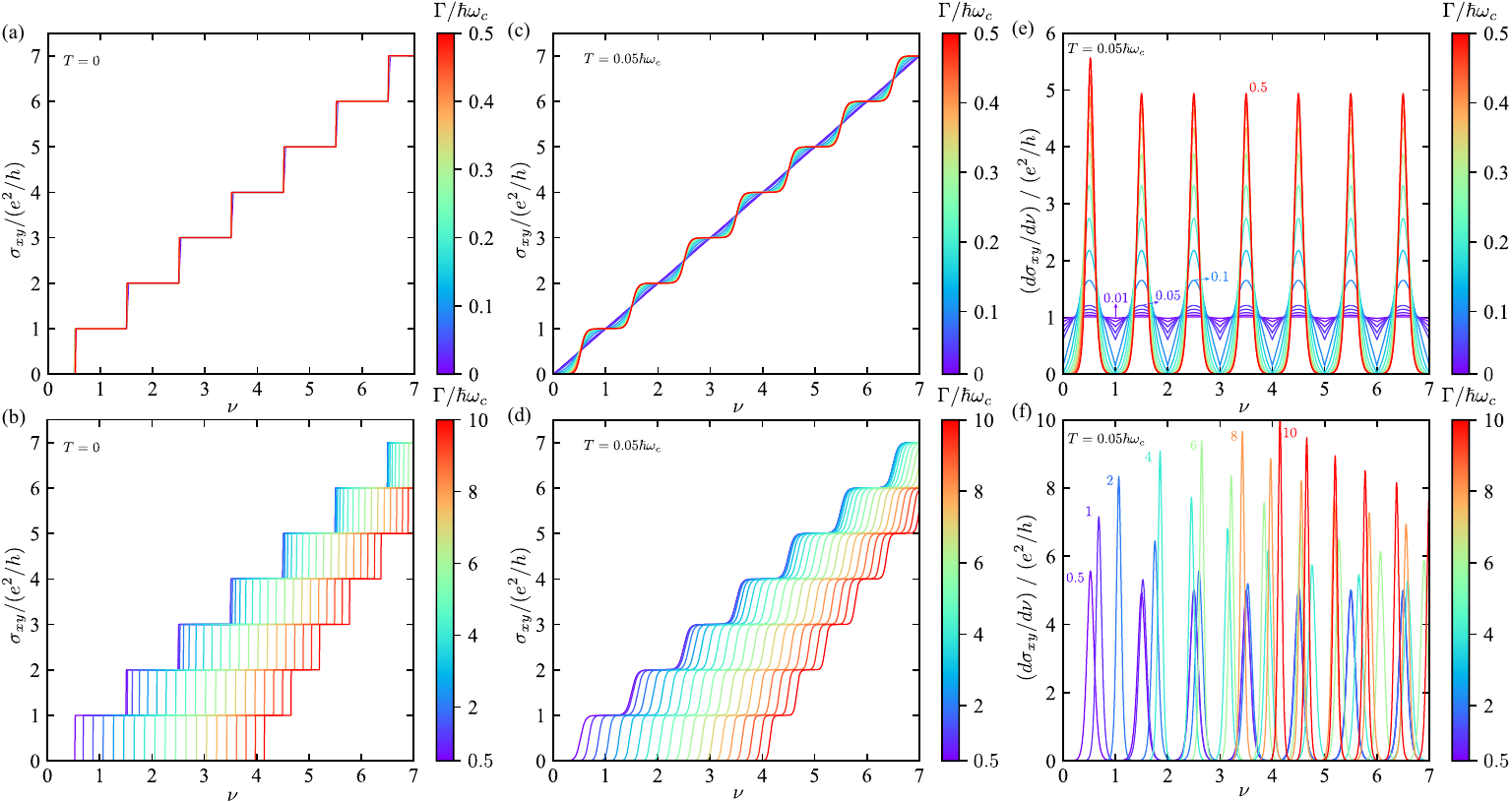}
    \caption{$T=0$ result of $\sigma_{xy}(\nu)$ for (a) weak disorder $\Gamma/\hbar\omega_c \in [0,0.5]$, and (b) strong disorder $\Gamma/\hbar\omega_c \in [0.5,10]$. 
    (c-d) are the corresponding results of $\sigma_{xy}(\nu)$ at a finite temperature $T/\hbar \omega_c=0.05$. (e-f) are the corresponding results of $d\sigma_{xy}/d\nu$ at a finite temperature $T/\hbar \omega_c=0.05$. The colored numbers label the value of $\Gamma/\hbar \omega_c$ used in the calculation.
    }
    \label{fig:sigmaxy_nu_7LL}
\end{figure*}

In this appendix section, we include the result of Hall conductivity of the percolation model for a larger range of disorder $\Gamma/\hbar \omega_c \in [0,10]$ and filling factors $\nu \in [0,7]$ shown in Fig.~\ref{fig:sigmaxy_nu_7LL} [cf.\ Fig.~~\ref{fig:sigmaxy_nu} in Section~\ref{sec:percolation_model}].
\end{document}